\batchmode
\makeatletter
\def\input@path{{C:/Users/51189/Documents/Data_IOP/TaS2/Manuscript/v7/}}
\makeatother
\documentclass[twocolumn,prl,showpacs,superscriptaddress,longbibliography]{revtex4-2}
\usepackage{textcomp}
\usepackage[latin9]{inputenc}
\usepackage{color}
\usepackage{amstext}
\usepackage{graphicx}
\usepackage{esint}
\usepackage[pdftex,pdfusetitle,
 bookmarks=true,bookmarksnumbered=false,bookmarksopen=false,
 breaklinks=false,pdfborder={0 0 0},pdfborderstyle={},backref=false,colorlinks=true]
 {hyperref}


\begin{document}
\title{Interlayer Hopping between Surface Mott Insulator and Bulk Band Insulator
in layered 1\emph{T}-TaS$_{2}$}
\author{Zijian Lin}
\thanks{Equally contributed to this work.}
\affiliation{International Center for Quantum Materials, School of Physics, Peking
University, Beijing 100871, People's Republic of China}
\affiliation{Beijing National Laboratory for Condensed Matter Physics and Institute
of Physics, Chinese Academy of Sciences, Beijing 100190, People's
Republic of China}
\author{Jie Li}
\thanks{Equally contributed to this work.}
\affiliation{National Laboratory of Solid State Microstructures and Department
of Physics, Nanjing University, Nanjing, 210093, People's Republic
of China}
\author{Xiaodong Cao}
\affiliation{Suzhou Institute for Advanced Research, University of Science and
Technology of China, Suzhou 215123, People's Republic of China}
\affiliation{School of Artificial Intelligence and Data Science, University of
Science and Technology of China, Suzhou 215123, People's Republic
of China}
\author{Jingjing Gao}
\affiliation{Key Laboratory of Materials Physics, Institute of Solid State Physics,
HFIPS, Chinese Academy of Sciences, Hefei 230031, People's Republic
of China}
\author{Xuan Luo}
\affiliation{Key Laboratory of Materials Physics, Institute of Solid State Physics,
HFIPS, Chinese Academy of Sciences, Hefei 230031, People's Republic
of China}
\author{Yuping Sun}
\email{ypsun@issp.ac.cn}

\affiliation{Anhui Key Laboratory of Low-Energy Quantum Materials and Devices,
High Magnetic Field Laboratory, HFIPS, Chinese Academy of Sciences,
Hefei 230031, People's Republic of China}
\affiliation{Key Laboratory of Materials Physics, Institute of Solid State Physics,
HFIPS, Chinese Academy of Sciences, Hefei 230031, People's Republic
of China}
\affiliation{Collaborative Innovation Center of Advanced Microstructures, Nanjing
University, Nanjing 210093, People's Republic of China}
\author{Yi Lu}
\email{yilu@nju.edu.cn}

\affiliation{National Laboratory of Solid State Microstructures and Department
of Physics, Nanjing University, Nanjing, 210093, People's Republic
of China}
\author{Nanlin Wang}
\email{nlwang@pku.edu.cn}

\affiliation{International Center for Quantum Materials, School of Physics, Peking
University, Beijing 100871, People's Republic of China}
\affiliation{Beijing Academy of Quantum Information Sciences, Beijing 100913, People's
Republic of China}
\author{Jiandong Guo}
\email{jdguo@iphy.ac.cn}

\affiliation{Beijing National Laboratory for Condensed Matter Physics and Institute
of Physics, Chinese Academy of Sciences, Beijing 100190, People's
Republic of China}
\affiliation{School of Physical Sciences, University of Chinese Academy of Sciences,
Beijing 100049, People's Republic of China}
\author{Xuetao Zhu}
\email{xtzhu@iphy.ac.cn}

\affiliation{Beijing National Laboratory for Condensed Matter Physics and Institute
of Physics, Chinese Academy of Sciences, Beijing 100190, People's
Republic of China}
\affiliation{School of Physical Sciences, University of Chinese Academy of Sciences,
Beijing 100049, People's Republic of China}
\begin{abstract}
In condensed matter physics, various mechanisms give rise to distinct
insulating phases. The competition and interplay between these phases
remain elusive, even for the seemingly most distinguishable band and
Mott insulators. In multilayer systems, such interplay is mediated
by interlayer hopping, which competes with the Coulomb repulsion to
determine the nature of insulators. The layered compound 1\emph{T}-TaS$_{2}$
provides an ideal platform for investigating this phenomenon, as it
naturally hosts coexisting Mott and band insulating states. However,
distinguishing these distinct insulating states and characterizing
the evolution remain challenging. In this study, we employ a dual
approach utilizing surface-sensitive High-Resolution Electron Energy
Loss Spectroscopy (HREELS) and bulk-sensitive Fourier-transform Infrared
Spectroscopy (FTIR) to investigate the electronic excitation spectrum
of 1\emph{T}-TaS$_{2}$. Our methodology effectively identifies the
features originating from the Mott and band insulators by analyzing
the differences in their bulk and surface spectral weights, along
with their energy distinctions. Based on the previous identification,
we further investigate the evolution of insulating state features
in the homostructure as they are modulated by temperature. The measurements
and Dynamical Mean-Field Theory (DMFT) calculations suggest that the
softening and broadening of Hubbard excitations in the Mott state
with increasing temperature result from enhanced interlayer hopping
between the Mott and band insulators.
\end{abstract}
\maketitle

\section{Introduction}

In condensed matter systems, various mechanisms can give rise to different
insulating phases, including band insulators, Mott insulators \citep{mottBasisElectronTheory1949,gebhardMottMetalInsulatorTransition1997},
charge/orbital ordered insulators \citep{tokuraOrbitalPhysicsTransitionMetal2000},
excitonic insulators \citep{jeromeExcitonicInsulator1967}, and Slater
insulators \citep{slaterMagneticEffectsHartreeFock1951}, among others.
While current research has focused on the emergence mechanisms of
exotic states, such as exciton condensation \citep{kogarSignaturesExcitonCondensation2017,linDramaticPlasmonResponse2022}
and metal-insulator transitions \citep{mottMetalInsulatorTransition1968},
a systematic understanding of the competition and interplay between
different insulating phases remains elusive, even for the seemingly
most distinguishable cases of band and Mott insulators. Band insulators,
which typically accommodate even numbers of electrons per unit cell,
are well described within the framework of Bloch's band theory, where
the periodic potential of the crystal lattice induces a gap between
the completely filled valence band and empty conduction band. In contrast,
Mott insulators, characterized by odd numbers of electron occupancy
per unit cell, manifest when the on-site Coulomb repulsion \emph{U}
exceeds a critical threshold relative to the electron kinetic energy
\citep{mottBasisElectronTheory1949,gebhardMottMetalInsulatorTransition1997}.
Although Mott insulators should exhibit metallic behavior according
to Bloch's band theory, strong electron correlations drive the splitting
of the metallic band at the Fermi level into upper and lower Hubbard
bands, thereby establishing an insulating state.

Multilayer systems with interlayer coupling provide a unique opportunity
to investigate the coupling between band and Mott insulators, due
to the competition between the Coulomb repulsion $U$ and interlayer
hopping $t_{\perp}$. For instance, within the bilayer Hubbard model,
a complex $\ensuremath{U}-\ensuremath{t_{\perp}}$ phase diagram,
which includes the Mott insulator, the correlated metal, and the band
insulator, was predicted \citep{capponiCurrentCarryingGround2004,kancharlaBandInsulatorMott2007,bouadimMagneticTransportProperties2008,rugerPhaseDiagramSquare2014,rademakerDeterminantQuantumMonte2013,golorGroundstatePhaseDiagram2014,leeCompetitionBandMott2014,najeraMultipleCrossoversCoherent2018}.
However, in a homogeneous system, the coexistence of Mott and band
insulators is typically prohibited, but the subtle differences in
environmental parameters between the surface and bulk may overcome
the limitation and cause the coexistence of different phases in systems
located at phase boundaries. Previous studies on artificial heterostructures
composed of Mott and band insulators \citep{ohtomoArtificialChargemodulationinAtomicscale2002,okamotoElectronicReconstructionInterface2004,okamotoTheoryMottInsulatorband2004,okamotoInterfaceOrderingPhase2005,seoOpticalStudyFreeCarrier2007}
have highlighted the critical importance of interfacial coupling.
Therefore, investigating and manipulating the coupling between different
surface and bulk phases, specifically through interlayer hopping $t_{\perp}$,
by varying environmental parameters presents a valuable opportunity
to deepen our understanding of the fundamental physics underlying
multilayer strongly correlated systems.

\begin{figure}
\noindent\begin{raggedright}
\includegraphics[width=8.6cm]{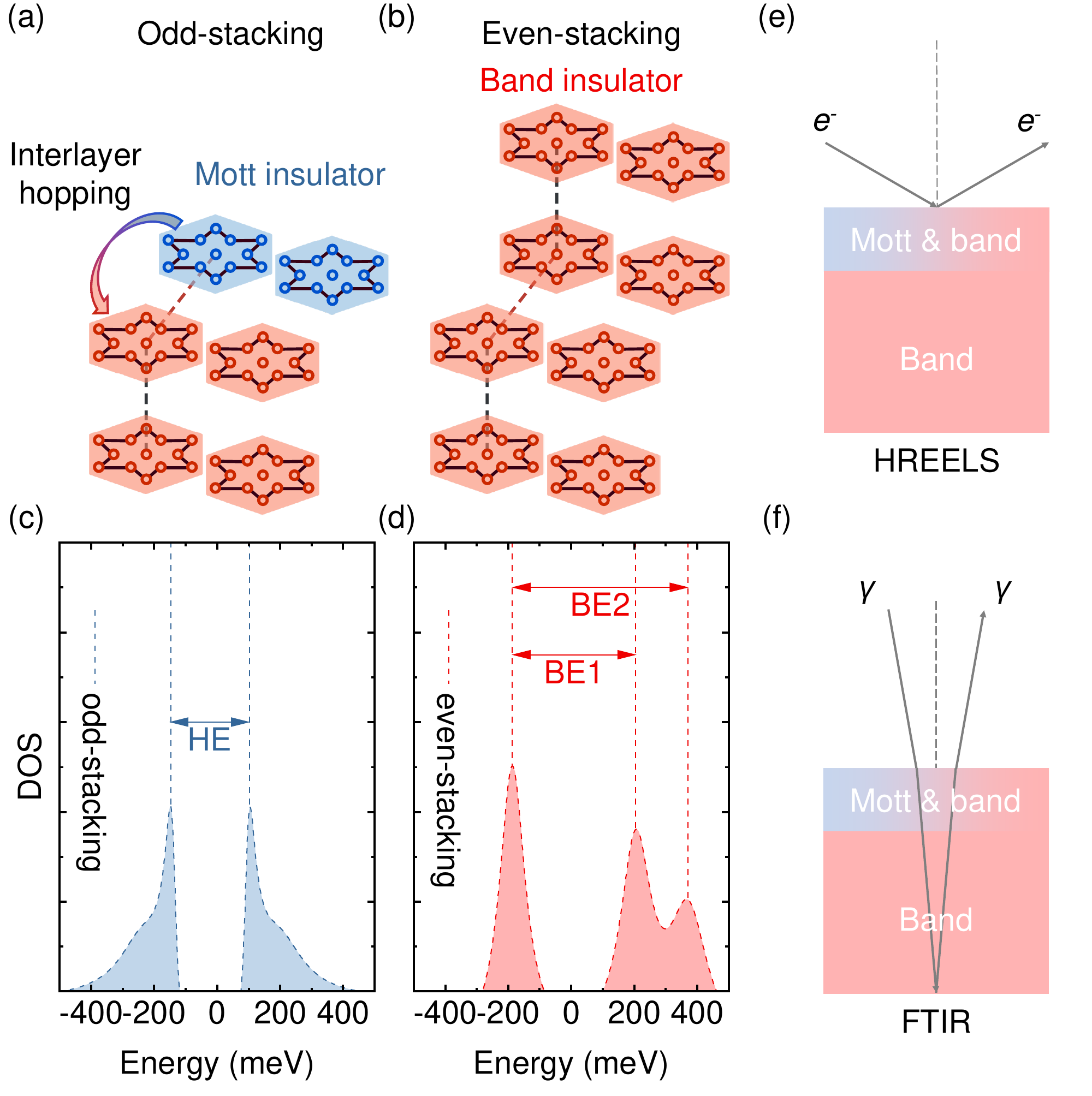}
\par\end{raggedright}
\caption{\protect\label{1}Comparison of the Mott and band insulator surface
in 1\emph{T}-TaS$_{2}$. (a) and (b) David star schematics in the
CCDW phase for different stacking. The odd-stacking corresponds to
the Mott surface, and the even-stacking corresponds to the band insulator
surface. (c) and (d) Density of states (DOS) schematics for odd- and
even-stacking. HE marks the Mott gap transition, and BE1 and BE2 denote
band insulator gap transitions. (e) HREELS probing surface features
($e^{-}$: electron beam). (f) FTIR probing bulk features ($\gamma$:
light beam).}
\end{figure}
A notable example of such complex interplay between different insulating
states can be found in 1\emph{T}-TaS$_{2}$. It is a strongly correlated
layered system where Mott and band insulators are believed to coexist.
The crystal structure of 1\emph{T}-TaS$_{2}$ is composed of S-Ta-S
trilayers, which are stacked sequentially via van der Waals interactions.
During cooling, it undergoes three charge density wave (CDW) transitions
at temperatures of about 550 K, 340 K, and 187 K, leading to four
distinct phases: the undistorted lattice, incommensurate charge density
wave (ICCDW), nearly commensurate charge density wave (NCCDW), and
commensurate charge density wave (CCDW) \citep{williamsDiffractionEvidenceKohn1974,scrubyRoleChargeDensity1975,fazekasElectricalStructuralMagnetic1979,thomsonScanningTunnelingMicroscopy1994,gasparovPhononAnomalyCharge2002,siposMottStateSuperconductivity2008,rossnagelOriginChargedensityWaves2011,wangBandInsulatorMott2020}.
A metal-insulator transition (MIT) occurs alongside the transition
from the metallic NCCDW phase to the insulating CCDW phase, with a
critical temperature during cooling at $T_{c}=187$ K. A hysteresis
occurs upon warming, with the temperature of MIT rising to $T_{w}=213$
K \citep{disalvoLowTemperatureElectrical1977}. In the insulator phase,
the distortion of the lattice is illustrated in Fig. \ref{1}(a) and
(b), where every cluster of 13 neighboring atoms forms a structure
known as a David star. Initially, the formation of David stars was
believed to be a Mott insulator \citep{fazekasElectricalStructuralMagnetic1979,kimObservationMottLocalization1994,siposMottStateSuperconductivity2008},
and then 1\emph{T}-TaS$_{2}$ is considered as one of the candidates
of the quantum spin liquid \citep{law1TTaS2Quantum2017,klanjsekHightemperatureQuantumSpin2017,heSpinonFermiSurface2018,wenExperimentalIdentificationQuantum2019},
due to the presence of an odd number of electrons with $S=1/2$ in
a huge David star. Meanwhile, recent studies \citep{ritschelOrbitalTexturesCharge2015,ritschelStackingdrivenGapFormation2018,leeOriginInsulatingPhase2019,butlerMottnessUnitcellDoubling2020,martinoPreferentialOutofplaneConduction2020}
show that interlayer dimerization (even numbers of electron for a
dimer) can result in the formation of band insulators in bulk of 1\emph{T}-TaS$_{2}$
{[}red David stars in Fig. \ref{1} (a) and (b){]}. In the case of
even-layer stacking {[}Fig. \ref{1}(b){]}, all layers form dimers,
causing the system completely to go into a band insulator state. In
contrast, for odd-layer stacking {[}Fig. \ref{1}(a){]}, a leaving
monolayer on the surface forms a single-layer Mott insulator \citep{butlerMottnessUnitcellDoubling2020,law1TTaS2Quantum2017,leeDistinguishingMottInsulator2021,petocchiMottHybridizationGap2022}.
Consequently, the odd-layer surface of 1\emph{T}-TaS$_{2}$ exhibits
the coexistence of Mott and band insulators. Therefore, the insulating
CCDW phase of 1\emph{T}-TaS$_{2}$ could serve as an ideal platform
for investigating how the Mottness evolves under the influence of
interlayer hopping $t_{\perp}$ between the surface Mott and bulk
insulators, in response to varying environmental parameters such as
temperature, strain \citep{nicholsonGapCollapseFlat2024}, or stacking
configuration \citep{wangDualisticInsulatorStates2024}.

However, the identification of Mott and band insulator gaps remains
controversial. On the one hand, bulk-sensitive studies \citep{stahlCollapseLayerDimerization2020,ramosSelectiveElectronPhonon2023}
have provided conclusive evidence of interlayer dimerization in the
bulk, leading some scanning tunneling spectroscopy (STS) studies \citep{thomsonScanningTunnelingMicroscopy1994,maMetallicMosaicPhase2016,leeDistinguishingMottInsulator2021,butlerMottnessUnitcellDoubling2020}
to attribute the smaller gap {[}Fig. \ref{1}(c){]} at the surface
to a Hubbard excitation (HE) of a Mott insulator and the larger gap
{[}Fig. \ref{1}(d){]} to band gap excitations (BE) in a band insulator,
based on the distinctions between odd and even stacking. On the other
hand, some research \citep{wuEffectStackingOrder2022,zhangReconcilingBulkMetallic2022,leeChargeDensityWave2023}
suggests that both the smaller and larger gaps at the surface are
characteristic of Mott insulators, showing insensitivity to surface
stacking. Furthermore, the temperature evolution of these gaps remains
unclear, due to the technical challenges in temperature control during
STS measurements and the inherent limitations of angle-resolved photoemission
spectroscopy (ARPES) in probing states above the Fermi level. To address
these challenges and provide additional insights into the gap identification
and temperature evolution in 1\emph{T}-TaS$_{2}$, complementary experimental
approaches are needed. High-resolution electron energy loss spectroscopy
(HREELS) and Fourier-transform infrared spectroscopy (FTIR) emerge
as particularly valuable techniques. Both methods can detect interband
excitations as manifested in the loss function $-\text{Im}[\varepsilon(\omega)^{-1}]$.
These techniques offer complementary spatial sensitivity: HREELS primarily
probes surface phenomena (with a detection depth \textless{} 5 nm
\citep{liaoPracticalElectronMicroscopy2006,politano3DDiracPlasmons2018})
{[}Fig. \ref{1}(e){]}, while FTIR accesses bulk properties (with
a detection depth \textgreater{} 1 $\mu$m for insulators) {[}Fig.
\ref{1}(f){]}. Additionally, both techniques facilitate reliable
temperature-dependent measurements, making them ideal tools for investigating
the evolution of both surface and bulk insulating states across different
temperature regimes.

Here, by combining surface-sensitive HREELS with bulk-sensitive FTIR,
we conducted a comprehensive study of the electronic excitation spectra
of 1\emph{T}-TaS$_{2}$. We observed several excitation features with
both techniques and identified the HE as a characteristic of Mott
insulators, the BE as indicative of band insulators, and the MIT-independent
interband excitations (IE) and the interband plasmons (IP) \citep{ghoshAnisotropicPlasmonsExcitons2017,jiaTopologicallyNontrivialInterband2020,hespObservationInterbandCollective2021}.
Comparing the spectral weight from both techniques, the DOS of previous
STS measurements \citep{thomsonScanningTunnelingMicroscopy1994,maMetallicMosaicPhase2016,leeDistinguishingMottInsulator2021,butlerMottnessUnitcellDoubling2020},
and dynamical mean-field theory (DMFT) calculations, we confirmed
that the Mott insulating state is confined to surface and subsurface
layers. Based on the previous identification, we further investigate
the evolution of insulating state features as they are modulated by
temperature. We discovered an anomalous softening of energy and broadening
in the width of the HE in the surface Mott insulator phase when approaching
the transition temperature. Further DMFT calculations suggest that
this behavior originates from enhanced interlayer hopping $t_{\perp}$
upon warming between the Mott insulating layer and the bulk band insulator.

\section{Method}

\subsection{Sample Growth}

\begin{figure*}[t]
\noindent\begin{raggedright}
\includegraphics[width=1\textwidth]{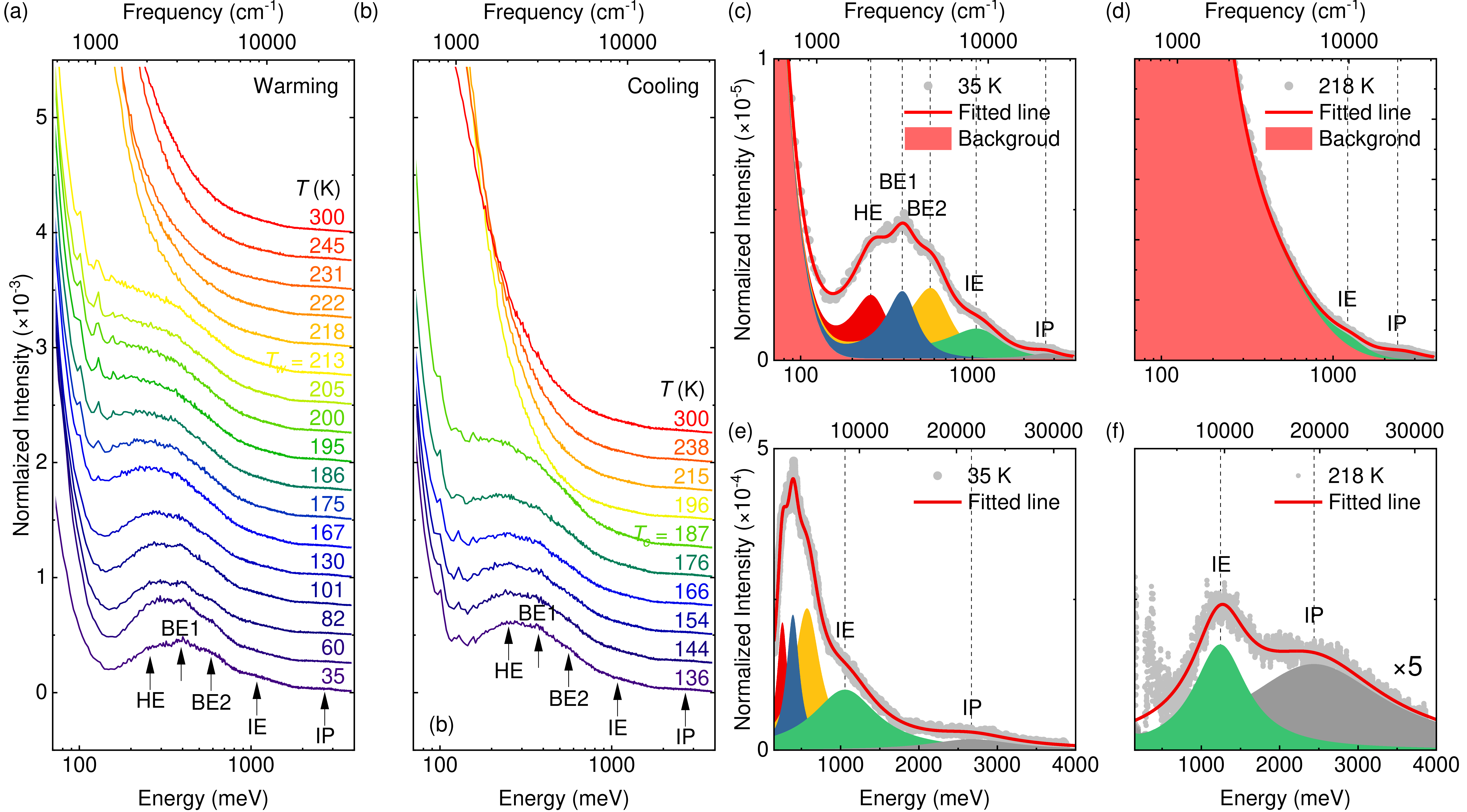}
\par\end{raggedright}
\caption{\protect\label{2}Temperature-dependent HREELS measurements of 1\emph{T}-TaS$_{2}$.
(a) and (b) HREELS measurements of warming and cooling processes,
respectively. The intensity is normalized to the elastic peak at zero
energy loss. The labeled peaks HE, BE, IE (interband excitation),
and IP (interband plasmon) are obtained from the fittings in (c) and
(d). (c) and (d) Fittings to the data measured at 35 K and 218 K during
the warming cycle. (e) and (f) display the data from (c) and (d) with
the background subtracted and a linear horizontal scale.}
\end{figure*}
High-quality 1\emph{T}-TaS$_{2}$ single crystals were synthesized
using the chemical vapor transport (CVT) method, employing iodine
as the transport medium. Tantalum (Ta, 99.99\%, sourced from Aladdin)
and sulfur (S, 99.99\%, also from Aladdin) powders were measured in
a 1:2 molar ratio and mixed with 0.2 g of iodine. This mixture was
then placed into silicon quartz tubes, which were subsequently sealed
under high vacuum conditions. The tubes were heated for 10 days in
a two-zone furnace with the source and growth zones set at temperatures
of 850 °C and 750 °C, respectively. After heating, the tubes were
quickly removed from the furnace and immediately quenched in an ice-water
mixture to cool rapidly, ultimately resulting in high-quality single-crystal
1\emph{T}-TaS$_{2}$.

\subsection{HREELS Measurement}

High-quality 1\emph{T}-TaS$_{2}$ crystals were mechanically cleaved
under ultra-high vacuum (1$\times$10$^{-10}$ mbar) for \emph{in
situ} measurements. The quality of the samples and the MIT temperatures
between NCCDW and CCDW phases, identified with $T_{c}=187$ K for
cooling and $T_{w}=213$ K for warming, were confirmed using Low-energy
electron diffraction (LEED) {[}see Supplementary Material (SM) for
further details \citep{SM}{]}. Subsequent HREELS measurements were
conducted with both the incident and scattering angles set to 65°,
using an incident electron energy of 110 eV. This incident energy
corresponds to the most surface-sensitive probe, with a detection
depth of only 5 Å \citep{liaoPracticalElectronMicroscopy2006,politano3DDiracPlasmons2018}.
Temperature-dependent measurements, including both warming and cooling
processes, were performed in a single experimental run. Details on
the principles and setup of the HREELS experiment are available in
the SM \citep{SM}.

\subsection{FTIR Measurement}

The FTIR measurements were conducted on cleaved hexagonal ab-plane
1\emph{T}-TaS$_{2}$ samples at ambient pressure. The near-normal
incidence setup allowed precise reflectance measurements over a frequency
range of 6 meV to 4 eV. To minimize stray light effects, samples were
mounted on a copper cone. Additionally, to ensure accurate reflectance
$R(\omega)$ data, an \emph{in situ} gold and aluminum evaporation
technique was applied at 300 K under a high vacuum of less than 5$\times$10$^{-7}$
mbar.

\subsection{DMFT Calculation}

Density functional theory (DFT) calculations were performed using
the all-electron, full-potential WIEN2K code with the augmented plane-wave
plus local orbital (APW+lo) basis set \citep{blahaWIEN2kAPW+loProgram2020}
and the Perdew-Burke-Ernzerhof (PBE) exchange functional \citep{perdewGeneralizedGradientApproximation1996a}.
The CCDW phase was obtained by optimizing both the lattice constants
and internal atomic coordinates of a $\sqrt{13}\times\sqrt{13}$ supercell
containing 13 Ta atoms, using a 10 $\times$ 10 two-dimensional \textbf{\emph{k}}-point
mesh. To derive an effective one-band tight-binding model \emph{H}$_{\text{mono}}$
for freestanding 1\emph{T}-TaS$_{2}$ monolayer, a maximally localized
Wannier orbital was constructed at the central Ta atom using the Wannier90
code \citep{mostofiWannier90ToolObtaining2008}. The bandwidth $w$
= 0.04 eV obtained from DFT calculations (see SM for details \citep{SM})
was employed in subsequent calculations of the insulating state.

To investigate the Mott insulating behavior of 1\emph{T}-TaS$_{2}$
in the CCDW phase, DMFT calculations were performed using the CTHyb
impurity solver \citep{sethTRIQSCTHYBContinuoustime2016}, as implemented
in the TRIQS library \citep{parcolletTRIQSToolboxResearch2015}. In
addition to the monolayer one-band calculation, interlayer hopping
$t_{\perp}$ in the slab model was considered by introducing intra-dimer
hopping between two layers of dimers {[}$t_{\text{aa}}$ in Fig. \ref{6}(a){]},
and inter-dimer hopping between Ta Wannier orbitals {[}$t_{\text{ab}}$
in Fig. \ref{6}(a){]}. $t_{\text{aa}}$ governs the bonding-antibonding
band splitting within a bilayer, which was set to 0.2 eV based on
experimental measurements of band excitation energies \citep{butlerMottnessUnitcellDoubling2020,leeDistinguishingMottInsulator2021}.
Due to the shift of the David stars between bilayers, $t_{\text{ab}}$
is expected to be significantly smaller than $t_{\text{aa}}$ and
is treated as an adjustable parameter in our calculations. For each
Ta orbital, the same non-zero $t_{\text{ab}}$ was included between
it and its two nearest neighbors in the adjacent layer.

The resulting Hamiltonian for a semi-infinite system with odd-stacking
{[}Fig. \ref{6}(a){]} \citep{leeOriginInsulatingPhase2019} assumes
a tridiagonal form. The Hamiltonian is truncated after seven layers
(one surface monolayer + three bilayer dimers), and the effect of
the rest of the system is included as an effective self-energy via
Löwdin downfolding, acting as an embedding potential on the 7th layer
\citep{petocchiMottHybridizationGap2022}, and expressed in a continued
fraction form:
\[
E(k_{\parallel},i\omega_{n})=\frac{|H_{\text{ab}}(k_{\parallel})|^{2}}{z_{6}-\frac{t_{\text{aa}}^{2}}{z_{7}-\frac{|H_{\text{ab}}(k_{\parallel})|^{2}}{z_{8}-\ldots}}}
\]
where $z_{i}=i\omega_{n}+\mu-H_{\text{mono}}(k_{\parallel})-\Sigma_{i}(k_{\parallel},i\omega_{n})$.
Here, $H_{\text{ab}}(k_{\parallel})$ represents the inter-dimer coupling
Hamiltonian in the momentum space, and $\Sigma_{i}(k_{\parallel},i\omega_{n})$
denotes formally the DMFT self-energy for each layer. We further assume
the depth-dependent self-energy is converged for $i\geq7$, which
is justified a posteriori by the calculation results. This seven-layer
slab model is subsequently solved in DMFT. The real-axis spectral
functions $A(\omega)$ are obtained using maximum entropy formalism
analytical continuation method \citep{krabergerMaximumEntropyFormalism2017},
and the optical conductivity is calculated as $\sigma(\Omega)\sim\int\text{d}\omega A\left(\omega+\frac{\Omega}{2}\right)A\left(\omega-\frac{\Omega}{2}\right)\frac{f\left(\omega-\frac{\Omega}{2}\right)-f\left(\omega+\frac{\Omega}{2}\right)}{\Omega}$,
where the $f$ is the Fermi-Dirac distribution and the effect of dipole
matrix elements is omitted.

\section{Result}

\subsection{HREELS}

Figures \ref{2}(a) and (b) present the results from HREELS measurements,
which clearly demonstrate significant spectral weight shifts associated
with the MIT, specifically the NCCDW to CCDW transition. The identified
MIT temperatures during the warming and cooling processes are approximately
$T_{w}=213$ K and $T_{c}=187$ K, respectively. These results align
with LEED observations and corroborate previous studies \citep{williamsDiffractionEvidenceKohn1974,scrubyRoleChargeDensity1975,gasparovPhononAnomalyCharge2002,wangBandInsulatorMott2020}.
Apart from the noticeable differences within the hysteresis temperature
range (187 K$<T<$213 K), the loss spectra during both the warming
and cooling processes remain consistent across other temperature ranges
(refer to the SM \citep{SM} for further data). Therefore, unless
otherwise specified, the following data and analysis are based on
the warming process. It is noteworthy that there is no evidence of
the intermediate Mott insulator phase, reported in previous research
\citep{wangBandInsulatorMott2020,dongDynamicsElectronicStates2023}
to occur between 213 K and 230 K during the warming phase.

To compare the differences between the metallic and insulating phases,
we applied a phenomenological fitting to the data (see fitting method
in the SM \citep{SM}) at 35 K and 218 K {[}Fig. \ref{2}(c) and (d){]}.
In the metallic NCCDW phase, the spectrum below 1000 meV shows a smooth,
continuous decline indicative of metallic and gapless behavior. In
contrast, in the insulating CCDW phase, a pronounced dip below 130
meV indicates an MIT. Furthermore, as temperature decreases, three
specific loss features, denoted as HE, BE1, and BE2, become progressively
more pronounced. Two consistent loss features, observed across all
experimental temperatures and labeled as IE and IP {[}Fig. \ref{2}(e)
and (f){]}, are considered unrelated to MIT. The detailed assignment
of these loss features will be discussed further later in the manuscript.

\begin{figure*}
\noindent\begin{raggedright}
\includegraphics[width=1\textwidth]{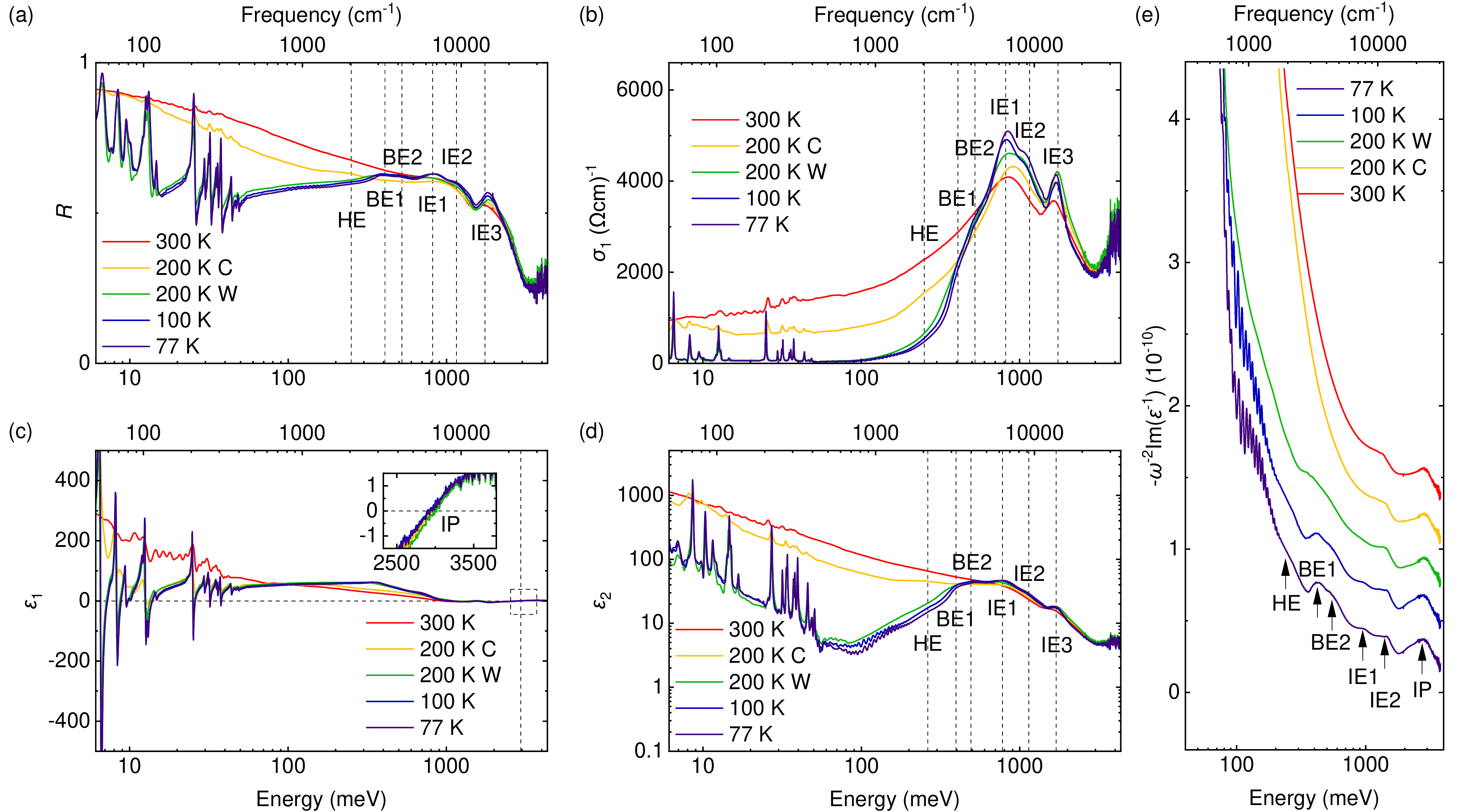}
\par\end{raggedright}
\caption{\protect\label{3}Temperature-dependent FTIR spectroscopy of 1\emph{T}-TaS$_{2}$.
(a) Reflectance of 1\emph{T}-TaS$_{2}$. (b), (c), (d), and (e) the
optical conductivity, the real part, the imaginary part of the dielectric
function, and the frequency-suppressed loss function, respectively.
All are derived from the Kramers-Kronig transformation applied to
the reflectance data in (a). Dashed lines in (a), (b), and (d) and
arrows in (e) highlight notable spectral features. In panel (c), the
dashed lines identify the zero-crossings of the real part of the dielectric
function, where it shifts from negative to positive values. The regions
enclosed by the dashed boxes are expanded into insets. The legends
indicate the temperatures at which the measurements were taken, with
'C' for the cooling cycle and 'W' for the warming cycle.}
\end{figure*}
HREELS is theoretically a convolution of the instrument's geometric
factor with the sample's loss function \citep{heinzExcitationPlasmonsInterband1980,gebhardMottMetalInsulatorTransition1997,lambinElectronenergylossSpectroscopyMultilayered1985,lambinComputationSurfaceElectronenergyloss1990,ibachElectronEnergyLoss2013,liGeometricEffectHighresolution2022a},
which is the imaginary part of the reciprocal of the dielectric function,
$-$Im$[\varepsilon(\omega)^{-1}]$ contributed from the sample surface.
On the other hand, FTIR can deduce the sample's dielectric function
$\varepsilon(\omega)$ of the bulk by measuring its optical reflectance
$R(\omega)$ and employing the Kramers-Kronig relations. Consequently,
by combining the insights gained from HREELS and FTIR measurements,
it becomes feasible to analyze the observed elementary excitations
and to separate the surface contributions thoroughly.

\subsection{FTIR}

Figure \ref{3}(a) displays the temperature-dependent FTIR measurement
results, covering a spectral range of up to 4 eV. Overall, the MIT
from the NCCDW to the CCDW phase significantly impacts the far- to
mid-infrared spectrum ($<$400 meV). In the high-temperature NCCDW
phase, the reflectance at the lowest measured energy of 6 meV is 0.95,
resembling that of a poor metal. The gradual decrease in reflectance
with increasing frequency lacks a clear plasma edge, consistent with
findings from earlier studies \citep{lucovskyReflectivityStudiesTi1976,deanPolaronicConductivityPhotoinduced2011,velebitEffectsSuperstructuringOptical2015}.
This pattern suggests significant damping of free carriers. Additionally,
the phonon mode intensity is weak, due to strong carrier screening
effects. In contrast, the low-temperature CCDW phase shows a significant
drop in low-frequency reflectance, indicative of insulating behavior,
along with a notable increase in phonon mode intensity due to reduced
electronic screening. Analyzing the fundamental spectral characteristics
at high and low temperatures reveals distinct spectral features at
200 K during both warming and cooling, corresponding to CCDW and NCCDW
phases, respectively, and demonstrating clear hysteresis and characteristics
of a first-order MIT.

In the high-temperature NCCDW phase and the low-temperature CCDW phase,
the Hagen-Rubens relation and constant low-energy extrapolation were
applied, respectively. The X-ray atomic scattering functions were
utilized for the high-energy extrapolation \citep{tannerUseXrayScattering2015}.
This approach enabled the determination of temperature-dependent optical
conductivity $\sigma_{1}(\omega)$ and both the real $\varepsilon_{1}(\omega)$
and imaginary parts $\varepsilon_{2}(\omega)$ of the dielectric function
via the Kramers-Kronig relations, as illustrated in Fig. \ref{3}(b)-(d).
In the CCDW phase at low temperatures, the $\sigma_{1}(\omega)$ at
low frequencies {[}Fig. \ref{3}(b){]} unveils zero conductivity along
with phonon features below the gap as a classic insulator system,
aligning with previous literature \citep{gasparovPhononAnomalyCharge2002,velebitEffectsSuperstructuringOptical2015}.
By analyzing the peaks in the $\sigma_{1}(\omega)$ and $\varepsilon_{2}(\omega)$
and combining the fitting results in the HREELS {[}Fig. \ref{2}(c){]},
six excitation features were identified: HE (there is almost no spectral
weight, which will be discussed in the next section), BE1, and BE2,
consistent with the HREELS results; IE1 and IE2, convoluted as IE
in the HREELS; and IE3, which generates IP in the HREELS and will
be discussed further later. The observed oscillatory features around
100 meV in the $\varepsilon_{2}(\omega)$, which can be attributed
to interference between the bottom and top surfaces of the sample
in the insulator phase, were methodically filtered and smoothed. This
interference indicates that light fully penetrates the sample, implying
that the bulk contributes the majority of the signal. In the high-temperature
NCCDW phase, the low-frequency $\sigma_{1}(\omega)$ deviates from
expected Drude behavior, manifesting as a plateau instead of a centered
peak at zero frequency. This behavior can be explained by effective
medium theory \citep{velebitEffectsSuperstructuringOptical2015},
which can be the origin of the substantial free carrier damping. With
an increase in temperature, excitations such as HE, BE1, and BE2 vanish,
IE1 and IE2 converge into a single feature, whereas IE3 remains largely
unaffected. Additionally, Fig. \ref{3}(c) highlights that the $\varepsilon_{1}(\omega)$
transitions from negative to positive at 2.9 eV, a shift typically
indicative of an interband plasmon \citep{ghoshAnisotropicPlasmonsExcitons2017,jiaTopologicallyNontrivialInterband2020,hespObservationInterbandCollective2021}.

\begin{figure}
\noindent\begin{raggedright}
\includegraphics[width=8.6cm]{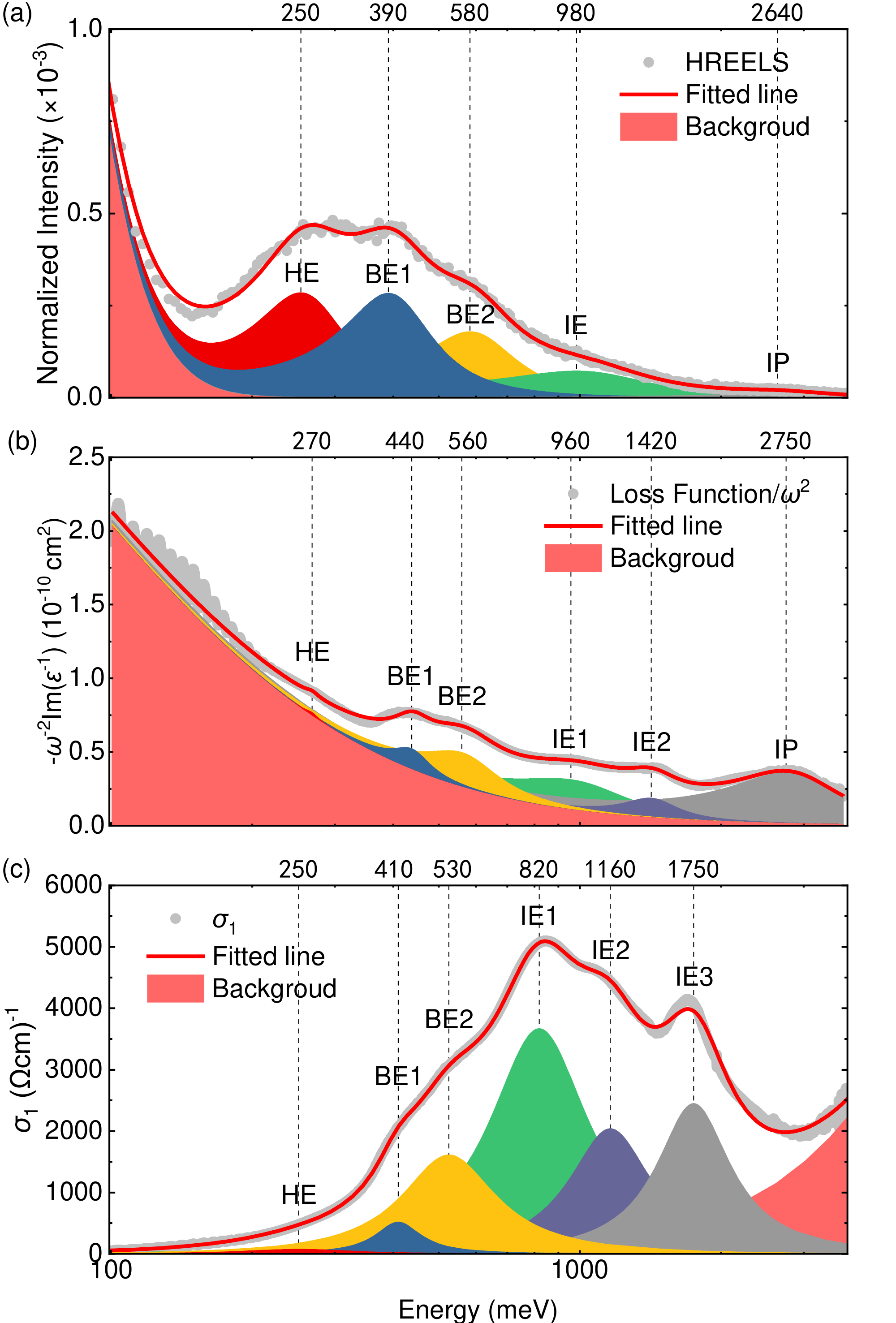}
\par\end{raggedright}
\caption{\protect\label{4}Comparison of HREELS and FTIR measurements in the
low-temperature NCCDW phase. (a), (b) and (c) The fitting results
for HREELS at 82 K, frequency-suppressed loss function and optical
conductivity measured by FTIR at 77 K, respectively.}
\end{figure}
To facilitate direct comparison with HREELS measurements, we derived
the frequency-suppressed loss function $-\omega^{-2}\text{Im}[\varepsilon(\omega)^{-1}]$
from $\varepsilon_{1}(\omega)$ and $\varepsilon_{2}(\omega)$ to
compare with HREELS results {[}Fig. \ref{3}(e){]}, where the frequency-suppressed
term $\omega^{-2}$ approximates the frequency-dependent component
of the HREELS geometric factor \citep{lambinComputationSurfaceElectronenergyloss1990}.
The frequency-suppressed loss function exhibits several characteristics
similar to those observed in HREELS, including the dip of around 130
meV in the CCDW phase and the presence of HE, BE1, BE2, IE1, and IP.
However, distinct differences are also observed, such as the relatively
weak spectral weight of HE, and the absence of IE2 in the HREELS results.

\section{Discussion}

\subsection{Distinguishing Hubbard Excitation from Band Insulator Excitation}

To quantitatively identify these electronic excitations, we applied
a phenomenological fitting (fit methods and result parameters seen
in the SM \citep{SM}) and compared the HREELS results at 82 K {[}Fig.
\ref{4}(a){]} and the frequency-suppressed loss function derived
from reflectance at 77 K {[}Fig. \ref{4}(b){]}. One can observe that,
except for IE2, the frequencies of all electronic excitations are
within a 10\% error margin between the two measurement methods. These
discrepancies could be attributed to the frequency suppression term
in the loss function, which only approximates the HREELS geometric
factor, leading to a significant influence of the background on the
fitting results. The most crucial quantitative difference lies in
the integrated spectral weight ratio of HE relative to the sum of
HE, BE1, and BE2, which is two orders of magnitude larger in HREELS
measurements (19\%) compared to FTIR (0.067\%). This quantitatively
suggests that HE excitations originate solely from surface contributions.

The physical attribution of these features remains unconfirmed, as
the loss function can capture both single-electron excitations (interband
transitions) and collective excitations (plasmons) \citep{ghoshAnisotropicPlasmonsExcitons2017,jiaTopologicallyNontrivialInterband2020,hespObservationInterbandCollective2021}.
Therefore, further comparison of the frequency-suppressed loss function
with optical conductivity, which only reflects single-electron excitations
in this energy range \footnote{Collective excitations related to atomic positions, such as phonons
and phase modes of charge density waves, can be measured at low energies.
However, common collective modes within the energy range we measured,
namely plasmons, are not reflected in the optical conductivity.}, is necessary to comprehend our results, with the Lorentz model usually
employed in semiconductor and insulator systems. The Lorentz model
can be expressed in the following form:

\[
\text{\ensuremath{\varepsilon(\omega)=\varepsilon_{\infty}+\sum_{i}\frac{(f_{i}\text{\ensuremath{\omega_{p}}})^{2}}{\omega_{i}^{2}-\omega^{2}-i\gamma_{i}\omega}}}
\]
where $\varepsilon_{\infty}$ is the high-frequency dielectric constant,
$\omega_{p}$ is the total plasma frequency contributed by all electronic
transitions, and $f_{i}$, $\omega_{i}$, $\gamma_{i}$ respectively
represent the fraction of the total number, the oscillator frequency,
and the damping coefficient of the $i^{th}$ single-electron excitation.

The contribution of a Lorentz oscillator to the optical conductivity
is characterized by a Lorentzian-like peak $\frac{\omega}{4\pi}\frac{(f_{i}\omega_{p})^{2}\gamma_{i}\omega}{\gamma_{i}^{2}\omega^{2}+(\omega_{i}^{2}-\omega^{2})^{2}}$
centered at the frequency $\omega_{i}$. The dashed lines in Fig.
\ref{4}(c) mark the $\omega_{i}$ of each Lorentz oscillator obtained
from the fitting. Compared to Fig. \ref{4}(b), the $\omega_{i}$
are consistently lower than the corresponding peak centers in Fig.
\ref{4}(c). This occurs because the peak center in Fig. \ref{4}(b)
is actually equal to $\sqrt{\frac{(f_{i}\omega_{p})^{2}}{\varepsilon_{\infty}}+\omega_{i}^{2}}$
(see details in the SM \citep{SM}). For $\frac{f_{i}\omega_{p}}{\sqrt{\varepsilon_{\infty}}}<<\ensuremath{\omega_{i}}$,
the difference in center frequencies is small, indicating that the
peak in the Fig. \ref{4}(b) comes from a single-particle excitation.
Conversely, a significant difference between peak centers suggests
the presence of a collective excitation, which also manifests in $\varepsilon_{1}(\omega)$
as a zero-crossing from negative to positive. Based on the above analysis,
the IP observed in the HREELS can be identified as an interband plasmon
\citep{ghoshAnisotropicPlasmonsExcitons2017,jiaTopologicallyNontrivialInterband2020,hespObservationInterbandCollective2021}
caused by IE3 because its peak center is much higher than that of
IE3 and it presents as a zero point in the $\varepsilon_{1}(\omega)$
{[}inset of the Fig. \ref{3}(c){]}; other features correspond to
single-electron excitations from the valence to the conduction band.

\begin{figure}
\includegraphics[width=8.6cm]{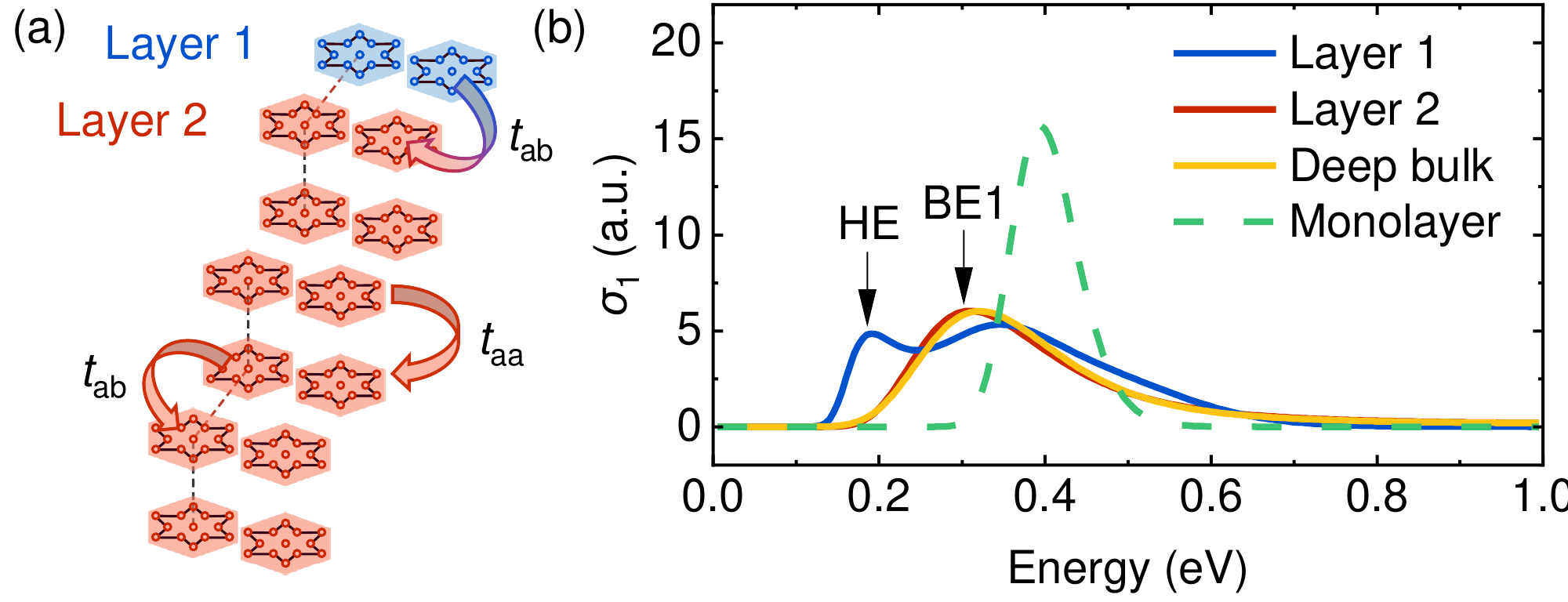}

\caption{\protect\label{6}Layer-resolved DMFT results at 30 K. (a) Schematic
of the seven-layer slab (one surface monolayer $+$ three bilayer
dimers) model used in the DMFT calculations. (b) Comparison of optical
conductivity results of the freestanding monolayer and the slab model,
where the deep bulk result is the analytic continuation of the seventh
layer.}
\end{figure}
 Furthermore, by comparing $\omega_{i}$ with the observed density
of states (DOS), we can determine the band origins of the HREELS features.
In our measurements, the energies corresponding to HE, BE1, and BE2
are 250, 390, and 580 meV, respectively. A previous STS study \citep{butlerMottnessUnitcellDoubling2020}
observed DOS for two types of stacking. As conceptually illustrated
in Fig. \ref{1}(c), for the odd-stacking, the highest valence band
(lower Hubbard band) is at -140 meV, and the lowest conduction band
(upper Hubbard band) is at 96 meV, with an energy difference of 236
meV; for even-stacking, the highest valence band is located at -187
meV, the lowest conduction band is at 205 meV, and the second-lowest
conduction band is at 372 meV, with differences of 392 and 559 meV,
respectively. Therefore, we can accurately assign HE to transitions
between upper and lower Hubbard bands in the Mott insulator and BE1
and BE2 to transitions between the highest valence and the lowest
conduction bands and the second-lowest conduction band in band insulators.

However, in contrast to this gap assignment, some recent studies \citep{wuEffectStackingOrder2022,zhangReconcilingBulkMetallic2022,leeChargeDensityWave2023}
have revealed that regardless of odd- or even-stacking, the majority
of the surface exhibits a large gap, while the small gap appears only
in minimal areas near step edges in odd-stacking. In our experiments,
however, the relative integrated spectral weight of 19\% for the small
gap (HE) indicates a contribution comparable in magnitude to that
of the large gap (BE). This discrepancy might be related to surface
reconstruction, where the Mott monolayer is buried at the third layer
rather than located at the topmost layer \citep{leeChargeDensityWave2023}.
Given that HREELS has a detection depth of \textless{} 5 nm, the relatively
large spectral weight could be attributed to subsurface contributions.
Due to the lack of definitive evidence, our subsequent discussion
will adhere to the scenario where the Mott monolayer is located at
the topmost surface.

To further support the identification of HE and BE1, we carried out
calculations of optical conductivity $\sigma_{1}$ within the DMFT
framework. The results for a freestanding monolayer are shown as the
green dashed line in Fig. \ref{6}(b). The DFT-derived tight-binding
Hamiltonian results in a single half-filled narrow Ta band at the
Fermi level with bandwidth $w\approx0.04$ eV (see SM for details
\citep{SM}). Using well-established parameters of $U=0.4$ eV \citep{yuElectronicCorrelationEffects2017,petocchiMottHybridizationGap2022}
at 30 K, the energy of the Hubbard gap transition was found to be
around 0.4 eV, which agrees with the previously reported monolayer
gap \citep{tianElectronicStructuresMott2024} but is significantly
higher than our experimentally observed surface value of 0.25 eV.
This discrepancy suggests that in multilayer 1\emph{T}-TaS$_{2}$,
the Mott state of the surface layer is non-trivially coupled to the
bulk system.

To effectively capture and simulate the influence of the bulk, we
adopted a seven-layer slab model (one surface monolayer $+$ three
bilayer dimers), in which the surface layer is coupled to the bulk
via inter-dimer hopping $t_{ab}$ {[}together with intra-dimer hopping
$t_{\text{aa}}$, they reflect the entire interlayer hopping $t_{\perp}$
shown as Fig. \ref{6}(a){]}.

\begin{figure}
\noindent\begin{raggedright}
\includegraphics[width=8.6cm]{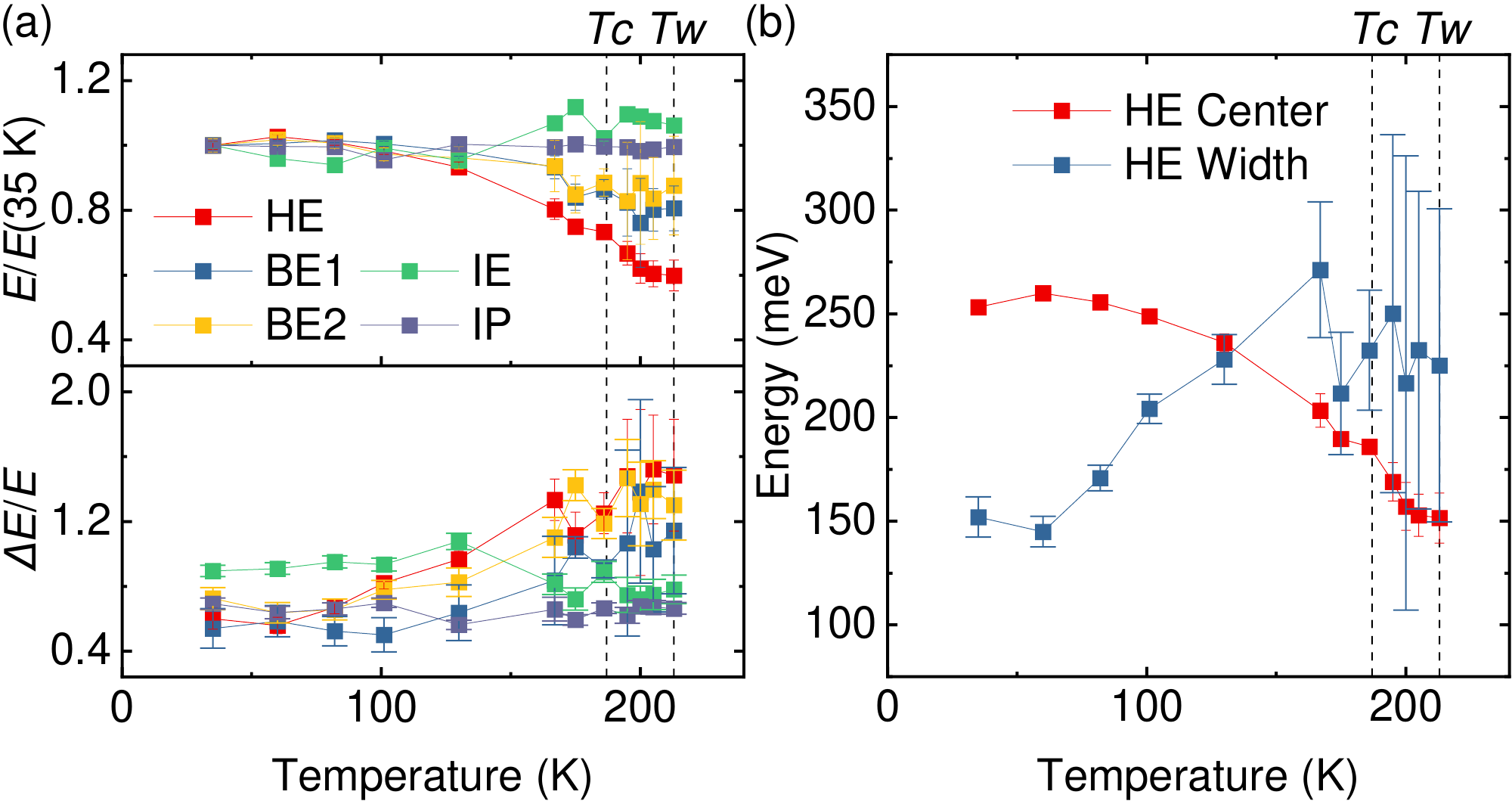}
\par\end{raggedright}
\caption{\protect\label{5}Temperature-dependent excitations in 1\emph{T}-TaS$_{2}$.
(a) Normalized energy and damping rate ($\Delta E/E$, where $\Delta E$
is the width) as determined from fitting the HREELS data in the warming
cycle. The energy has been normalized to the measurements at 35 K
for comparison. $T_{c}$ and $T_{w}$ mark the MIT temperatures for
the cooling cycle and the warming cycle, respectively. (c) Temperature
evolution of the peak center and width of the HE.}
\end{figure}
Layer-resolved results {[}Fig. \ref{6}(b){]} demonstrate that with
the same parameters ($U=0.4$ eV), introducing $t_{ab}=0.05$ eV effectively
lowers the energy of HE, bringing it into good agreement with the
experimental data. Moreover, the energies of Layer 2 and the deep
bulk layers are substantially higher than that of HE, confirming that
the lowest-energy excitation is indeed surface HE, while the higher-energy
excitations correspond to bulk BE.

\subsection{Temperature-dependent Hubbard excitation modulated by interlayer
hopping}

Our temperature-dependent results further emphasize the critical role
of the interlayer hopping $t_{ab}$ in the behavior of 1\emph{T}-TaS$_{2}$.
Figure \ref{5}(a) presents the temperature dependence of electronic
excitations during the warming process. The energies and widths of
the MIT-independent excitations, IE and IP, remain almost constant
upon warming. BE1 and BE2 show very subtle temperature dependence.
In contrast, HE shows significant softening and broadening as the
temperature rises. Notably, above 130 K, the broadening of HE exceeds
the energy at its peak {[}Fig. \ref{5}(b){]}, suggesting that the
Hubbard band gap is nearly closed.

\begin{figure}
\includegraphics[width=8.6cm]{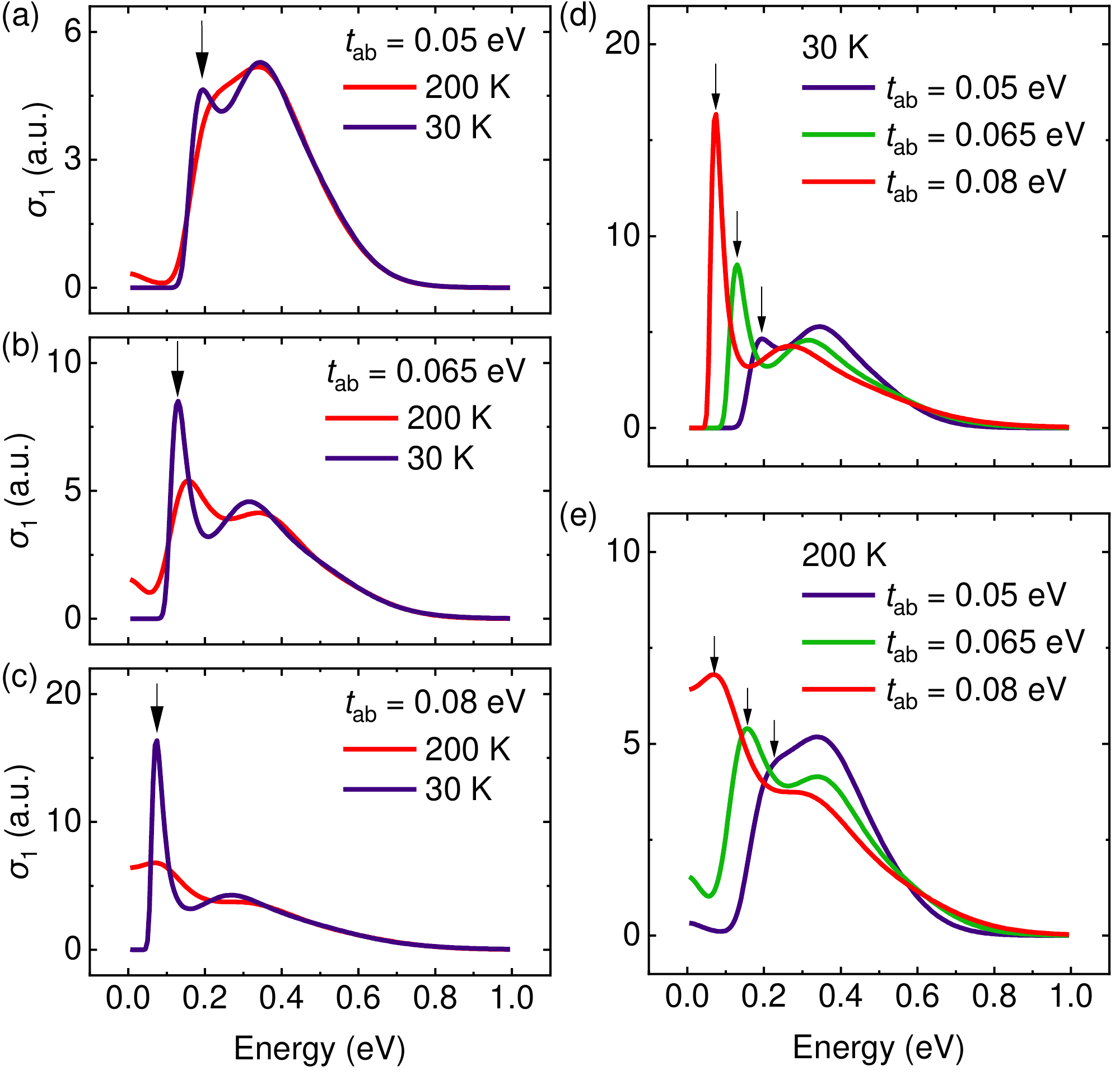}

\caption{\protect\label{7} Modulated Hubbard excitations by interlayer hopping
and temperature. (a) Optical conductivity of Layer 1 at 200 K and
30 K calculated within DMFT framework by slab model, with the same
parameters as in Fig. \ref{6}(b). Arrows indicate features arising
from interband transitions from the lower to upper Hubbard bands.
(b) and (c) Same as panel (a) with $t_{ab}=0.065$ and 0.08 eV, respectively.
(d) and (e) Same data as in (a)--(c) but plotted together at the
same temperature for easier comparison of the effects of $t_{ab}.$}
\end{figure}

The Mott insulator gap is determined by the Coulomb repulsion $U$.
Generally, a decrease in $U$ leads to a smaller Hubbard gap, with
more pronounced broadening as the temperature increases (see SM \citep{SM}).
However, $U$ typically depends on the element and remains nearly
constant with temperature. Indeed, our DMFT calculations confirm that
using the same parameters as in Fig. \ref{6}(b) ($U=0.4$ eV, and
$t_{ab}=0.05$ eV) to simulate the results at 200 K {[}Fig. \ref{7}(a){]}
fails to capture the experimentally observed energy reduction and
broadening. Therefore, the significant changes in HE in Fig. \ref{5}(b)
must be influenced by factors beyond just the temperature effect.

The discussion in Fig. \ref{6} and earlier theoretical studies \citep{capponiCurrentCarryingGround2004,kancharlaBandInsulatorMott2007,bouadimMagneticTransportProperties2008,rugerPhaseDiagramSquare2014,rademakerDeterminantQuantumMonte2013,golorGroundstatePhaseDiagram2014,leeCompetitionBandMott2014,najeraMultipleCrossoversCoherent2018}
suggest that interlayer hopping plays a significant role in modulating
the Hubbard gap. Figure \ref{7}(c) shows that increasing $t_{ab}$
significantly reduces the Hubbard gap excitation at 30 K, but meanwhile
decreases the broadening. Only when the thermal effect is considered
can the calculation successfully lead to a broadening of the HE {[}Fig.
\ref{7}(b) and (c){]}. The maximum broadening is observed at $t_{ab}=0.08$
eV at 200 K, where the Hubbard gap even closes entirely. The combined
influence of $t_{ab}$ and thermal effects {[}Fig. \ref{7}(e){]}
further explains the behavior that HE exhibits both energy reduction
and broadening simultaneously. Therefore, our calculations indicate
that the temperature evolution of HE is jointly driven by $t_{ab}$
and thermal effects. More specifically, the gradual rise in $t_{ab}$
modulates HE to lower energies as the temperature increases, while
the combination of $t_{ab}$ and thermal effects contributes to the
increased broadening.

We propose two possible mechanisms that could lead to the observed
increase of $t_{ab}$ as temperature increases: interlayer sliding
of the David star and a decrease in interlayer distance. In the first
scenario, some experiments \citep{maMetallicMosaicPhase2016,nicholsonGapCollapseFlat2024,wangDualisticInsulatorStates2024}
have demonstrated the shift of HE energy, attributed to different
sites of surface sliding. Additionally, electric or optical field-induced
metastable states \citep{stojchevskaUltrafastSwitchingStable2014,ritschelOrbitalTexturesCharge2015,gerasimenkoIntertwinedChiralCharge2019,maObservationMultipleMetastable2019},
caused by sliding, have been observed, which might explain the more
pronounced reduction of HE energy in the warming hysteresis region.
The second scenario, involving a decrease in the interlayer distance,
seems to contradict the classical behavior of thermal expansion. However,
in 1\emph{T}-TaS$_{2}$, the interlayer distance of the surface Mott
layer (approximately 7 Å) is significantly larger than that of the
bulk layers (around 6 Å) \citep{givensThermalExpansionOp1977,butlerMottnessUnitcellDoubling2020,yangOriginDistinctInsulating2024},
which deviates from the typical surface contraction due to surface
tension. Hence, the temperature dependence of the interlayer distance
in the surface Mott layer is expected to be complex. Both of these
possibilities merit more detailed experiments using variable-temperature
surface probes.

\section{Conclusion}

In this study, we systematically investigated the coexistence and
coupling between Mott and band insulator phases in 1\emph{T}-TaS$_{2}$,
focusing particularly on how interlayer hopping modulates the surface
Mott state. By combining surface-sensitive HREELS with bulk-sensitive
FTIR spectroscopy, we achieved two major findings:

First, we definitively identified and distinguished the electronic
excitations characteristic of different insulating states. The Hubbard
excitation at 250 meV represents the Mott insulator phase, while band
excitations (BE1 at 390 meV and BE2 at 580 meV) correspond to the
band insulator phase. These assignments are strongly supported by
both the consistency with previous STS measurements \citep{thomsonScanningTunnelingMicroscopy1994,maMetallicMosaicPhase2016,leeDistinguishingMottInsulator2021,butlerMottnessUnitcellDoubling2020}
and our DMFT calculations. Most importantly, the dramatic contrast
in HE spectral weight between HREELS (19\%) and FTIR (0.067\%) measurements
provides compelling evidence that the Mott insulating state is confined
to the surface and subsurface layers, while the bulk maintains a band
insulator state.

Second, we revealed the crucial role of interlayer hopping in modulating
the surface Mott state. Our temperature-dependent measurements demonstrated
that the HE feature exhibits significant softening and broadening
upon warming toward the MIT. Through detailed DMFT calculations, we
showed that this behavior originates from the combined effect of enhanced
interlayer hopping ($t_{\text{ab}}$) and thermal effect. This finding
highlights how the coupling between surface Mott and bulk band insulator
states can be tuned through environmental parameters.
\begin{acknowledgments}
The HREELS measurements presented in this work were conducted at the
second affiliation, and the FTIR measurements were performed at the
first affiliation. This work was supported by the National Key R\&D
Program of China (No. 2022YFA1403000 and No. 2021YFA1400200), the
National Natural Science Foundation of China (No. 12274446), and the
Strategic Priority Research Program of Chinese Academy of Sciences
(No. XDB33000000). Yi Lu was supported by National Natural Science
Foundation of China (No. 12274207). Nanlin Wang was supported by National
Natural Science Foundation of China (No. 12488201), and the National
Key R\&D Program of China (No. 2022YFA1403901). Jingjing Gao, Xuan
Luo, and Yuping Sun were supported by the National Key R\&D Program
(No. 2023YFA1607402 and No. 2021YFA1600201), the National Natural
Science Foundation of China (No. U2032215, No. U1932217, and No. 12274412),
and Systematic Fundamental Research Program Leveraging Major Scientific
and Technological Infrastructure, Chinese Academy of Sciences under
Contract No. JZHKYPT- 437 2021-08.
\end{acknowledgments}

\bibliography{7C__Users_51189_Documents_Data_IOP_TaS2_Manuscript_v7_ref,8C__Users_51189_Documents_Data_IOP_TaS2_Manuscript_v7_SM}

\begin{thebibliography}{82}%
\makeatletter
\providecommand \@ifxundefined [1]{%
 \@ifx{#1\undefined}
}%
\providecommand \@ifnum [1]{%
 \ifnum #1\expandafter \@firstoftwo
 \else \expandafter \@secondoftwo
 \fi
}%
\providecommand \@ifx [1]{%
 \ifx #1\expandafter \@firstoftwo
 \else \expandafter \@secondoftwo
 \fi
}%
\providecommand \natexlab [1]{#1}%
\providecommand \enquote  [1]{``#1''}%
\providecommand \bibnamefont  [1]{#1}%
\providecommand \bibfnamefont [1]{#1}%
\providecommand \citenamefont [1]{#1}%
\providecommand \href@noop [0]{\@secondoftwo}%
\providecommand \href [0]{\begingroup \@sanitize@url \@href}%
\providecommand \@href[1]{\@@startlink{#1}\@@href}%
\providecommand \@@href[1]{\endgroup#1\@@endlink}%
\providecommand \@sanitize@url [0]{\catcode `\\12\catcode `\$12\catcode
  `\&12\catcode `\#12\catcode `\^12\catcode `\_12\catcode `\%12\relax}%
\providecommand \@@startlink[1]{}%
\providecommand \@@endlink[0]{}%
\providecommand \url  [0]{\begingroup\@sanitize@url \@url }%
\providecommand \@url [1]{\endgroup\@href {#1}{\urlprefix }}%
\providecommand \urlprefix  [0]{URL }%
\providecommand \Eprint [0]{\href }%
\providecommand \doibase [0]{https://doi.org/}%
\providecommand \selectlanguage [0]{\@gobble}%
\providecommand \bibinfo  [0]{\@secondoftwo}%
\providecommand \bibfield  [0]{\@secondoftwo}%
\providecommand \translation [1]{[#1]}%
\providecommand \BibitemOpen [0]{}%
\providecommand \bibitemStop [0]{}%
\providecommand \bibitemNoStop [0]{.\EOS\space}%
\providecommand \EOS [0]{\spacefactor3000\relax}%
\providecommand \BibitemShut  [1]{\csname bibitem#1\endcsname}%
\let\auto@bib@innerbib\@empty
\bibitem [{\citenamefont {Mott}(1949)}]{mottBasisElectronTheory1949}%
  \BibitemOpen
  \bibfield  {author} {\bibinfo {author} {\bibfnamefont {N.~F.}\ \bibnamefont
  {Mott}},\ }\bibfield  {title} {\bibinfo {title} {The {{Basis}} of the
  {{Electron Theory}} of {{Metals}}, with {{Special Reference}} to the
  {{Transition Metals}}},\ }\href {https://doi.org/10.1088/0370-1298/62/7/303}
  {\bibfield  {journal} {\bibinfo  {journal} {Proc. Phys. Soc. A}\ }\textbf
  {\bibinfo {volume} {62}},\ \bibinfo {pages} {416} (\bibinfo {year}
  {1949})}\BibitemShut {NoStop}%
\bibitem [{\citenamefont
  {Gebhard}(1997)}]{gebhardMottMetalInsulatorTransition1997}%
  \BibitemOpen
  \bibfield  {author} {\bibinfo {author} {\bibfnamefont {F.}~\bibnamefont
  {Gebhard}},\ }\href {https://doi.org/10.1007/3-540-14858-2} {\emph {\bibinfo
  {title} {The {{Mott Metal-Insulator Transition}}: {{Models}} and
  {{Methods}}}}},\ \bibinfo {series} {Springer {{Tracts}} in {{Modern
  Physics}}}, Vol.\ \bibinfo {volume} {137}\ (\bibinfo  {publisher}
  {Springer},\ \bibinfo {address} {Berlin, Heidelberg},\ \bibinfo {year}
  {1997})\BibitemShut {NoStop}%
\bibitem [{\citenamefont {Tokura}\ and\ \citenamefont
  {Nagaosa}(2000)}]{tokuraOrbitalPhysicsTransitionMetal2000}%
  \BibitemOpen
  \bibfield  {author} {\bibinfo {author} {\bibfnamefont {Y.}~\bibnamefont
  {Tokura}}\ and\ \bibinfo {author} {\bibfnamefont {N.}~\bibnamefont
  {Nagaosa}},\ }\bibfield  {title} {\bibinfo {title} {Orbital {{Physics}} in
  {{Transition-Metal Oxides}}},\ }\href
  {https://doi.org/10.1126/science.288.5465.462} {\bibfield  {journal}
  {\bibinfo  {journal} {Science}\ }\textbf {\bibinfo {volume} {288}},\ \bibinfo
  {pages} {462} (\bibinfo {year} {2000})}\BibitemShut {NoStop}%
\bibitem [{\citenamefont {J{\'e}rome}\ \emph {et~al.}(1967)\citenamefont
  {J{\'e}rome}, \citenamefont {Rice},\ and\ \citenamefont
  {Kohn}}]{jeromeExcitonicInsulator1967}%
  \BibitemOpen
  \bibfield  {author} {\bibinfo {author} {\bibfnamefont {D.}~\bibnamefont
  {J{\'e}rome}}, \bibinfo {author} {\bibfnamefont {T.~M.}\ \bibnamefont
  {Rice}},\ and\ \bibinfo {author} {\bibfnamefont {W.}~\bibnamefont {Kohn}},\
  }\bibfield  {title} {\bibinfo {title} {Excitonic {{Insulator}}},\ }\href
  {https://doi.org/10.1103/PhysRev.158.462} {\bibfield  {journal} {\bibinfo
  {journal} {Phys. Rev.}\ }\textbf {\bibinfo {volume} {158}},\ \bibinfo {pages}
  {462} (\bibinfo {year} {1967})}\BibitemShut {NoStop}%
\bibitem [{\citenamefont
  {Slater}(1951)}]{slaterMagneticEffectsHartreeFock1951}%
  \BibitemOpen
  \bibfield  {author} {\bibinfo {author} {\bibfnamefont {J.~C.}\ \bibnamefont
  {Slater}},\ }\bibfield  {title} {\bibinfo {title} {Magnetic {{Effects}} and
  the {{Hartree-Fock Equation}}},\ }\href
  {https://doi.org/10.1103/PhysRev.82.538} {\bibfield  {journal} {\bibinfo
  {journal} {Phys. Rev.}\ }\textbf {\bibinfo {volume} {82}},\ \bibinfo {pages}
  {538} (\bibinfo {year} {1951})}\BibitemShut {NoStop}%
\bibitem [{\citenamefont {Kogar}\ \emph {et~al.}(2017)\citenamefont {Kogar},
  \citenamefont {Rak}, \citenamefont {Vig}, \citenamefont {Husain},
  \citenamefont {Flicker}, \citenamefont {Joe}, \citenamefont {Venema},
  \citenamefont {MacDougall}, \citenamefont {Chiang}, \citenamefont {Fradkin},
  \citenamefont {{van Wezel}},\ and\ \citenamefont
  {Abbamonte}}]{kogarSignaturesExcitonCondensation2017}%
  \BibitemOpen
  \bibfield  {author} {\bibinfo {author} {\bibfnamefont {A.}~\bibnamefont
  {Kogar}}, \bibinfo {author} {\bibfnamefont {M.~S.}\ \bibnamefont {Rak}},
  \bibinfo {author} {\bibfnamefont {S.}~\bibnamefont {Vig}}, \bibinfo {author}
  {\bibfnamefont {A.~A.}\ \bibnamefont {Husain}}, \bibinfo {author}
  {\bibfnamefont {F.}~\bibnamefont {Flicker}}, \bibinfo {author} {\bibfnamefont
  {Y.~I.}\ \bibnamefont {Joe}}, \bibinfo {author} {\bibfnamefont
  {L.}~\bibnamefont {Venema}}, \bibinfo {author} {\bibfnamefont {G.~J.}\
  \bibnamefont {MacDougall}}, \bibinfo {author} {\bibfnamefont {T.~C.}\
  \bibnamefont {Chiang}}, \bibinfo {author} {\bibfnamefont {E.}~\bibnamefont
  {Fradkin}}, \bibinfo {author} {\bibfnamefont {J.}~\bibnamefont {{van
  Wezel}}},\ and\ \bibinfo {author} {\bibfnamefont {P.}~\bibnamefont
  {Abbamonte}},\ }\bibfield  {title} {\bibinfo {title} {Signatures of exciton
  condensation in a transition metal dichalcogenide},\ }\href
  {https://doi.org/10.1126/science.aam6432} {\bibfield  {journal} {\bibinfo
  {journal} {Science}\ }\textbf {\bibinfo {volume} {358}},\ \bibinfo {pages}
  {1314} (\bibinfo {year} {2017})}\BibitemShut {NoStop}%
\bibitem [{\citenamefont {Lin}\ \emph {et~al.}(2022)\citenamefont {Lin},
  \citenamefont {Wang}, \citenamefont {Balassis}, \citenamefont {Echeverry},
  \citenamefont {Vasenko}, \citenamefont {Silkin}, \citenamefont {Chulkov},
  \citenamefont {Shi}, \citenamefont {Zhang}, \citenamefont {Guo},\ and\
  \citenamefont {Zhu}}]{linDramaticPlasmonResponse2022}%
  \BibitemOpen
  \bibfield  {author} {\bibinfo {author} {\bibfnamefont {Z.}~\bibnamefont
  {Lin}}, \bibinfo {author} {\bibfnamefont {C.}~\bibnamefont {Wang}}, \bibinfo
  {author} {\bibfnamefont {A.}~\bibnamefont {Balassis}}, \bibinfo {author}
  {\bibfnamefont {J.~P.}\ \bibnamefont {Echeverry}}, \bibinfo {author}
  {\bibfnamefont {A.~S.}\ \bibnamefont {Vasenko}}, \bibinfo {author}
  {\bibfnamefont {V.~M.}\ \bibnamefont {Silkin}}, \bibinfo {author}
  {\bibfnamefont {E.~V.}\ \bibnamefont {Chulkov}}, \bibinfo {author}
  {\bibfnamefont {Y.}~\bibnamefont {Shi}}, \bibinfo {author} {\bibfnamefont
  {J.}~\bibnamefont {Zhang}}, \bibinfo {author} {\bibfnamefont
  {J.}~\bibnamefont {Guo}},\ and\ \bibinfo {author} {\bibfnamefont
  {X.}~\bibnamefont {Zhu}},\ }\bibfield  {title} {\bibinfo {title} {Dramatic
  {{Plasmon Response}} to the {{Charge-Density-Wave Gap Development}} in
  1{{{\emph{T}}}}-{{TiSe}}{\textsubscript{2}}},\ }\href
  {https://doi.org/10.1103/PhysRevLett.129.187601} {\bibfield  {journal}
  {\bibinfo  {journal} {Phys. Rev. Lett.}\ }\textbf {\bibinfo {volume} {129}},\
  \bibinfo {pages} {187601} (\bibinfo {year} {2022})}\BibitemShut {NoStop}%
\bibitem [{\citenamefont {MOTT}(1968)}]{mottMetalInsulatorTransition1968}%
  \BibitemOpen
  \bibfield  {author} {\bibinfo {author} {\bibfnamefont {N.~F.}\ \bibnamefont
  {MOTT}},\ }\bibfield  {title} {\bibinfo {title} {Metal-{{Insulator
  Transition}}},\ }\href {https://doi.org/10.1103/RevModPhys.40.677} {\bibfield
   {journal} {\bibinfo  {journal} {Rev. Mod. Phys.}\ }\textbf {\bibinfo
  {volume} {40}},\ \bibinfo {pages} {677} (\bibinfo {year} {1968})}\BibitemShut
  {NoStop}%
\bibitem [{\citenamefont {Capponi}\ \emph {et~al.}(2004)\citenamefont
  {Capponi}, \citenamefont {Wu},\ and\ \citenamefont
  {Zhang}}]{capponiCurrentCarryingGround2004}%
  \BibitemOpen
  \bibfield  {author} {\bibinfo {author} {\bibfnamefont {S.}~\bibnamefont
  {Capponi}}, \bibinfo {author} {\bibfnamefont {C.}~\bibnamefont {Wu}},\ and\
  \bibinfo {author} {\bibfnamefont {S.-C.}\ \bibnamefont {Zhang}},\ }\bibfield
  {title} {\bibinfo {title} {Current carrying ground state in a bilayer model
  of strongly correlated systems},\ }\href
  {https://doi.org/10.1103/PhysRevB.70.220505} {\bibfield  {journal} {\bibinfo
  {journal} {Phys. Rev. B}\ }\textbf {\bibinfo {volume} {70}},\ \bibinfo
  {pages} {220505} (\bibinfo {year} {2004})}\BibitemShut {NoStop}%
\bibitem [{\citenamefont {Kancharla}\ and\ \citenamefont
  {Okamoto}(2007)}]{kancharlaBandInsulatorMott2007}%
  \BibitemOpen
  \bibfield  {author} {\bibinfo {author} {\bibfnamefont {S.~S.}\ \bibnamefont
  {Kancharla}}\ and\ \bibinfo {author} {\bibfnamefont {S.}~\bibnamefont
  {Okamoto}},\ }\bibfield  {title} {\bibinfo {title} {Band insulator to
  {{Mott}} insulator transition in a bilayer {{Hubbard}} model},\ }\href
  {https://doi.org/10.1103/PhysRevB.75.193103} {\bibfield  {journal} {\bibinfo
  {journal} {Phys. Rev. B}\ }\textbf {\bibinfo {volume} {75}},\ \bibinfo
  {pages} {193103} (\bibinfo {year} {2007})}\BibitemShut {NoStop}%
\bibitem [{\citenamefont {Bouadim}\ \emph {et~al.}(2008)\citenamefont
  {Bouadim}, \citenamefont {Batrouni}, \citenamefont {H{\'e}bert},\ and\
  \citenamefont {Scalettar}}]{bouadimMagneticTransportProperties2008}%
  \BibitemOpen
  \bibfield  {author} {\bibinfo {author} {\bibfnamefont {K.}~\bibnamefont
  {Bouadim}}, \bibinfo {author} {\bibfnamefont {G.~G.}\ \bibnamefont
  {Batrouni}}, \bibinfo {author} {\bibfnamefont {F.}~\bibnamefont
  {H{\'e}bert}},\ and\ \bibinfo {author} {\bibfnamefont {R.~T.}\ \bibnamefont
  {Scalettar}},\ }\bibfield  {title} {\bibinfo {title} {Magnetic and transport
  properties of a coupled {{Hubbard}} bilayer with electron and hole doping},\
  }\href {https://doi.org/10.1103/PhysRevB.77.144527} {\bibfield  {journal}
  {\bibinfo  {journal} {Phys. Rev. B}\ }\textbf {\bibinfo {volume} {77}},\
  \bibinfo {pages} {144527} (\bibinfo {year} {2008})}\BibitemShut {NoStop}%
\bibitem [{\citenamefont {R{\"u}ger}\ \emph {et~al.}(2014)\citenamefont
  {R{\"u}ger}, \citenamefont {Tocchio}, \citenamefont {Valent{\'i}},\ and\
  \citenamefont {Gros}}]{rugerPhaseDiagramSquare2014}%
  \BibitemOpen
  \bibfield  {author} {\bibinfo {author} {\bibfnamefont {R.}~\bibnamefont
  {R{\"u}ger}}, \bibinfo {author} {\bibfnamefont {L.~F.}\ \bibnamefont
  {Tocchio}}, \bibinfo {author} {\bibfnamefont {R.}~\bibnamefont
  {Valent{\'i}}},\ and\ \bibinfo {author} {\bibfnamefont {C.}~\bibnamefont
  {Gros}},\ }\bibfield  {title} {\bibinfo {title} {The phase diagram of the
  square lattice bilayer {{Hubbard}} model: A variational {{Monte Carlo}}
  study},\ }\href {https://doi.org/10.1088/1367-2630/16/3/033010} {\bibfield
  {journal} {\bibinfo  {journal} {New J. Phys.}\ }\textbf {\bibinfo {volume}
  {16}},\ \bibinfo {pages} {033010} (\bibinfo {year} {2014})}\BibitemShut
  {NoStop}%
\bibitem [{\citenamefont {Rademaker}\ \emph {et~al.}(2013)\citenamefont
  {Rademaker}, \citenamefont {Johnston}, \citenamefont {Zaanen},\ and\
  \citenamefont {{van den Brink}}}]{rademakerDeterminantQuantumMonte2013}%
  \BibitemOpen
  \bibfield  {author} {\bibinfo {author} {\bibfnamefont {L.}~\bibnamefont
  {Rademaker}}, \bibinfo {author} {\bibfnamefont {S.}~\bibnamefont {Johnston}},
  \bibinfo {author} {\bibfnamefont {J.}~\bibnamefont {Zaanen}},\ and\ \bibinfo
  {author} {\bibfnamefont {J.}~\bibnamefont {{van den Brink}}},\ }\bibfield
  {title} {\bibinfo {title} {Determinant quantum {{Monte Carlo}} study of
  exciton condensation in the bilayer {{Hubbard}} model},\ }\href
  {https://doi.org/10.1103/PhysRevB.88.235115} {\bibfield  {journal} {\bibinfo
  {journal} {Phys. Rev. B}\ }\textbf {\bibinfo {volume} {88}},\ \bibinfo
  {pages} {235115} (\bibinfo {year} {2013})}\BibitemShut {NoStop}%
\bibitem [{\citenamefont {Golor}\ \emph {et~al.}(2014)\citenamefont {Golor},
  \citenamefont {Reckling}, \citenamefont {Classen}, \citenamefont {Scherer},\
  and\ \citenamefont {Wessel}}]{golorGroundstatePhaseDiagram2014}%
  \BibitemOpen
  \bibfield  {author} {\bibinfo {author} {\bibfnamefont {M.}~\bibnamefont
  {Golor}}, \bibinfo {author} {\bibfnamefont {T.}~\bibnamefont {Reckling}},
  \bibinfo {author} {\bibfnamefont {L.}~\bibnamefont {Classen}}, \bibinfo
  {author} {\bibfnamefont {M.~M.}\ \bibnamefont {Scherer}},\ and\ \bibinfo
  {author} {\bibfnamefont {S.}~\bibnamefont {Wessel}},\ }\bibfield  {title}
  {\bibinfo {title} {Ground-state phase diagram of the half-filled bilayer
  {{Hubbard}} model},\ }\href {https://doi.org/10.1103/PhysRevB.90.195131}
  {\bibfield  {journal} {\bibinfo  {journal} {Phys. Rev. B}\ }\textbf {\bibinfo
  {volume} {90}},\ \bibinfo {pages} {195131} (\bibinfo {year}
  {2014})}\BibitemShut {NoStop}%
\bibitem [{\citenamefont {Lee}\ \emph {et~al.}(2014)\citenamefont {Lee},
  \citenamefont {Zhang}, \citenamefont {Jeschke},\ and\ \citenamefont
  {Valent{\'i}}}]{leeCompetitionBandMott2014}%
  \BibitemOpen
  \bibfield  {author} {\bibinfo {author} {\bibfnamefont {H.}~\bibnamefont
  {Lee}}, \bibinfo {author} {\bibfnamefont {Y.-Z.}\ \bibnamefont {Zhang}},
  \bibinfo {author} {\bibfnamefont {H.~O.}\ \bibnamefont {Jeschke}},\ and\
  \bibinfo {author} {\bibfnamefont {R.}~\bibnamefont {Valent{\'i}}},\
  }\bibfield  {title} {\bibinfo {title} {Competition between band and {{Mott}}
  insulators in the bilayer {{Hubbard}} model: {{A}} dynamical cluster
  approximation study},\ }\href {https://doi.org/10.1103/PhysRevB.89.035139}
  {\bibfield  {journal} {\bibinfo  {journal} {Phys. Rev. B}\ }\textbf {\bibinfo
  {volume} {89}},\ \bibinfo {pages} {035139} (\bibinfo {year}
  {2014})}\BibitemShut {NoStop}%
\bibitem [{\citenamefont {N{\'a}jera}\ \emph {et~al.}(2018)\citenamefont
  {N{\'a}jera}, \citenamefont {Civelli}, \citenamefont {Dobrosavljevi{\'c}},\
  and\ \citenamefont {Rozenberg}}]{najeraMultipleCrossoversCoherent2018}%
  \BibitemOpen
  \bibfield  {author} {\bibinfo {author} {\bibfnamefont {O.}~\bibnamefont
  {N{\'a}jera}}, \bibinfo {author} {\bibfnamefont {M.}~\bibnamefont {Civelli}},
  \bibinfo {author} {\bibfnamefont {V.}~\bibnamefont {Dobrosavljevi{\'c}}},\
  and\ \bibinfo {author} {\bibfnamefont {M.~J.}\ \bibnamefont {Rozenberg}},\
  }\bibfield  {title} {\bibinfo {title} {Multiple crossovers and coherent
  states in a {{Mott-Peierls}} insulator},\ }\href
  {https://doi.org/10.1103/PhysRevB.97.045108} {\bibfield  {journal} {\bibinfo
  {journal} {Phys. Rev. B}\ }\textbf {\bibinfo {volume} {97}},\ \bibinfo
  {pages} {045108} (\bibinfo {year} {2018})}\BibitemShut {NoStop}%
\bibitem [{\citenamefont {Ohtomo}\ \emph {et~al.}(2002)\citenamefont {Ohtomo},
  \citenamefont {Muller}, \citenamefont {Grazul},\ and\ \citenamefont
  {Hwang}}]{ohtomoArtificialChargemodulationinAtomicscale2002}%
  \BibitemOpen
  \bibfield  {author} {\bibinfo {author} {\bibfnamefont {A.}~\bibnamefont
  {Ohtomo}}, \bibinfo {author} {\bibfnamefont {D.~A.}\ \bibnamefont {Muller}},
  \bibinfo {author} {\bibfnamefont {J.~L.}\ \bibnamefont {Grazul}},\ and\
  \bibinfo {author} {\bibfnamefont {H.~Y.}\ \bibnamefont {Hwang}},\ }\bibfield
  {title} {\bibinfo {title} {Artificial charge-modulationin atomic-scale
  perovskite titanate superlattices},\ }\href
  {https://doi.org/10.1038/nature00977} {\bibfield  {journal} {\bibinfo
  {journal} {Nature}\ }\textbf {\bibinfo {volume} {419}},\ \bibinfo {pages}
  {378} (\bibinfo {year} {2002})}\BibitemShut {NoStop}%
\bibitem [{\citenamefont {Okamoto}\ and\ \citenamefont
  {Millis}(2004{\natexlab{a}})}]{okamotoElectronicReconstructionInterface2004}%
  \BibitemOpen
  \bibfield  {author} {\bibinfo {author} {\bibfnamefont {S.}~\bibnamefont
  {Okamoto}}\ and\ \bibinfo {author} {\bibfnamefont {A.~J.}\ \bibnamefont
  {Millis}},\ }\bibfield  {title} {\bibinfo {title} {Electronic reconstruction
  at an interface between a {{Mott}} insulator and a band insulator},\ }\href
  {https://doi.org/10.1038/nature02450} {\bibfield  {journal} {\bibinfo
  {journal} {Nature}\ }\textbf {\bibinfo {volume} {428}},\ \bibinfo {pages}
  {630} (\bibinfo {year} {2004}{\natexlab{a}})}\BibitemShut {NoStop}%
\bibitem [{\citenamefont {Okamoto}\ and\ \citenamefont
  {Millis}(2004{\natexlab{b}})}]{okamotoTheoryMottInsulatorband2004}%
  \BibitemOpen
  \bibfield  {author} {\bibinfo {author} {\bibfnamefont {S.}~\bibnamefont
  {Okamoto}}\ and\ \bibinfo {author} {\bibfnamefont {A.~J.}\ \bibnamefont
  {Millis}},\ }\bibfield  {title} {\bibinfo {title} {Theory of {{Mott}}
  insulator--band insulator heterostructures},\ }\href
  {https://doi.org/10.1103/PhysRevB.70.075101} {\bibfield  {journal} {\bibinfo
  {journal} {Phys. Rev. B}\ }\textbf {\bibinfo {volume} {70}},\ \bibinfo
  {pages} {075101} (\bibinfo {year} {2004}{\natexlab{b}})}\BibitemShut
  {NoStop}%
\bibitem [{\citenamefont {Okamoto}\ and\ \citenamefont
  {Millis}(2005)}]{okamotoInterfaceOrderingPhase2005}%
  \BibitemOpen
  \bibfield  {author} {\bibinfo {author} {\bibfnamefont {S.}~\bibnamefont
  {Okamoto}}\ and\ \bibinfo {author} {\bibfnamefont {A.~J.}\ \bibnamefont
  {Millis}},\ }\bibfield  {title} {\bibinfo {title} {Interface ordering and
  phase competition in a model {{Mott-insulator-band-insulator}}
  heterostructure},\ }\href {https://doi.org/10.1103/PhysRevB.72.235108}
  {\bibfield  {journal} {\bibinfo  {journal} {Phys. Rev. B}\ }\textbf {\bibinfo
  {volume} {72}},\ \bibinfo {pages} {235108} (\bibinfo {year}
  {2005})}\BibitemShut {NoStop}%
\bibitem [{\citenamefont {Seo}\ \emph {et~al.}(2007)\citenamefont {Seo},
  \citenamefont {Choi}, \citenamefont {Lee}, \citenamefont {Yu}, \citenamefont
  {Kim}, \citenamefont {Bernhard},\ and\ \citenamefont
  {Noh}}]{seoOpticalStudyFreeCarrier2007}%
  \BibitemOpen
  \bibfield  {author} {\bibinfo {author} {\bibfnamefont {S.~S.~A.}\
  \bibnamefont {Seo}}, \bibinfo {author} {\bibfnamefont {W.~S.}\ \bibnamefont
  {Choi}}, \bibinfo {author} {\bibfnamefont {H.~N.}\ \bibnamefont {Lee}},
  \bibinfo {author} {\bibfnamefont {L.}~\bibnamefont {Yu}}, \bibinfo {author}
  {\bibfnamefont {K.~W.}\ \bibnamefont {Kim}}, \bibinfo {author} {\bibfnamefont
  {C.}~\bibnamefont {Bernhard}},\ and\ \bibinfo {author} {\bibfnamefont
  {T.~W.}\ \bibnamefont {Noh}},\ }\bibfield  {title} {\bibinfo {title} {Optical
  {{Study}} of the {{Free-Carrier Response}} of {{LaTiO}}$_3$/{{SrTiO}}$_3$
  {{Superlattices}}},\ }\href {https://doi.org/10.1103/PhysRevLett.99.266801}
  {\bibfield  {journal} {\bibinfo  {journal} {Phys. Rev. Lett.}\ }\textbf
  {\bibinfo {volume} {99}},\ \bibinfo {pages} {266801} (\bibinfo {year}
  {2007})}\BibitemShut {NoStop}%
\bibitem [{\citenamefont {Williams}\ \emph {et~al.}(1974)\citenamefont
  {Williams}, \citenamefont {Parry},\ and\ \citenamefont
  {Scrub}}]{williamsDiffractionEvidenceKohn1974}%
  \BibitemOpen
  \bibfield  {author} {\bibinfo {author} {\bibfnamefont {P.~M.}\ \bibnamefont
  {Williams}}, \bibinfo {author} {\bibfnamefont {G.~S.}\ \bibnamefont
  {Parry}},\ and\ \bibinfo {author} {\bibfnamefont {C.~B.}\ \bibnamefont
  {Scrub}},\ }\bibfield  {title} {\bibinfo {title} {Diffraction evidence for
  the {{Kohn}} anomaly in {{1$T$-TaS$_2$}}},\ }\href
  {https://doi.org/10.1080/14786437408213248} {\bibfield  {journal} {\bibinfo
  {journal} {Philos. Mag.}\ }\textbf {\bibinfo {volume} {29}},\ \bibinfo
  {pages} {695} (\bibinfo {year} {1974})}\BibitemShut {NoStop}%
\bibitem [{\citenamefont {Scruby}\ \emph {et~al.}(1975)\citenamefont {Scruby},
  \citenamefont {Williams},\ and\ \citenamefont
  {Parry}}]{scrubyRoleChargeDensity1975}%
  \BibitemOpen
  \bibfield  {author} {\bibinfo {author} {\bibfnamefont {C.~B.}\ \bibnamefont
  {Scruby}}, \bibinfo {author} {\bibfnamefont {P.~M.}\ \bibnamefont
  {Williams}},\ and\ \bibinfo {author} {\bibfnamefont {G.~S.}\ \bibnamefont
  {Parry}},\ }\bibfield  {title} {\bibinfo {title} {The role of charge density
  waves in structural transformations of {{1$T$-TaS$_2$}}},\ }\href
  {https://doi.org/10.1080/14786437508228930} {\bibfield  {journal} {\bibinfo
  {journal} {Philos. Mag.}\ }\textbf {\bibinfo {volume} {31}},\ \bibinfo
  {pages} {255} (\bibinfo {year} {1975})}\BibitemShut {NoStop}%
\bibitem [{\citenamefont {Fazekas}\ and\ \citenamefont
  {Tosatti}(1979)}]{fazekasElectricalStructuralMagnetic1979}%
  \BibitemOpen
  \bibfield  {author} {\bibinfo {author} {\bibfnamefont {P.}~\bibnamefont
  {Fazekas}}\ and\ \bibinfo {author} {\bibfnamefont {E.}~\bibnamefont
  {Tosatti}},\ }\bibfield  {title} {\bibinfo {title} {Electrical, structural
  and magnetic properties of pure and doped {{1$T$-TaS$_2$}}},\ }\href
  {https://doi.org/10.1080/13642817908245359} {\bibfield  {journal} {\bibinfo
  {journal} {Philos. Mag. B}\ }\textbf {\bibinfo {volume} {39}},\ \bibinfo
  {pages} {229} (\bibinfo {year} {1979})}\BibitemShut {NoStop}%
\bibitem [{\citenamefont {Thomson}\ \emph {et~al.}(1994)\citenamefont
  {Thomson}, \citenamefont {Burk}, \citenamefont {Zettl},\ and\ \citenamefont
  {Clarke}}]{thomsonScanningTunnelingMicroscopy1994}%
  \BibitemOpen
  \bibfield  {author} {\bibinfo {author} {\bibfnamefont {R.~E.}\ \bibnamefont
  {Thomson}}, \bibinfo {author} {\bibfnamefont {B.}~\bibnamefont {Burk}},
  \bibinfo {author} {\bibfnamefont {A.}~\bibnamefont {Zettl}},\ and\ \bibinfo
  {author} {\bibfnamefont {J.}~\bibnamefont {Clarke}},\ }\bibfield  {title}
  {\bibinfo {title} {Scanning tunneling microscopy of the charge-density-wave
  structure in {{1$T$-TaS$_2$}}},\ }\href
  {https://doi.org/10.1103/PhysRevB.49.16899} {\bibfield  {journal} {\bibinfo
  {journal} {Phys. Rev. B}\ }\textbf {\bibinfo {volume} {49}},\ \bibinfo
  {pages} {16899} (\bibinfo {year} {1994})}\BibitemShut {NoStop}%
\bibitem [{\citenamefont {Gasparov}\ \emph {et~al.}(2002)\citenamefont
  {Gasparov}, \citenamefont {Brown}, \citenamefont {Wint}, \citenamefont
  {Tanner}, \citenamefont {Berger}, \citenamefont {Margaritondo}, \citenamefont
  {Ga{\'a}l},\ and\ \citenamefont
  {Forr{\'o}}}]{gasparovPhononAnomalyCharge2002}%
  \BibitemOpen
  \bibfield  {author} {\bibinfo {author} {\bibfnamefont {L.~V.}\ \bibnamefont
  {Gasparov}}, \bibinfo {author} {\bibfnamefont {K.~G.}\ \bibnamefont {Brown}},
  \bibinfo {author} {\bibfnamefont {A.~C.}\ \bibnamefont {Wint}}, \bibinfo
  {author} {\bibfnamefont {D.~B.}\ \bibnamefont {Tanner}}, \bibinfo {author}
  {\bibfnamefont {H.}~\bibnamefont {Berger}}, \bibinfo {author} {\bibfnamefont
  {G.}~\bibnamefont {Margaritondo}}, \bibinfo {author} {\bibfnamefont
  {R.}~\bibnamefont {Ga{\'a}l}},\ and\ \bibinfo {author} {\bibfnamefont
  {L.}~\bibnamefont {Forr{\'o}}},\ }\bibfield  {title} {\bibinfo {title}
  {Phonon anomaly at the charge ordering transition in
  1{{{\emph{T}}}}-{{TaS}}{\textsubscript{2}}},\ }\href
  {https://doi.org/10.1103/PhysRevB.66.094301} {\bibfield  {journal} {\bibinfo
  {journal} {Phys. Rev. B}\ }\textbf {\bibinfo {volume} {66}},\ \bibinfo
  {pages} {094301} (\bibinfo {year} {2002})}\BibitemShut {NoStop}%
\bibitem [{\citenamefont {Sipos}\ \emph {et~al.}(2008)\citenamefont {Sipos},
  \citenamefont {Kusmartseva}, \citenamefont {Akrap}, \citenamefont {Berger},
  \citenamefont {Forr{\'o}},\ and\ \citenamefont {Tuti{\v
  s}}}]{siposMottStateSuperconductivity2008}%
  \BibitemOpen
  \bibfield  {author} {\bibinfo {author} {\bibfnamefont {B.}~\bibnamefont
  {Sipos}}, \bibinfo {author} {\bibfnamefont {A.~F.}\ \bibnamefont
  {Kusmartseva}}, \bibinfo {author} {\bibfnamefont {A.}~\bibnamefont {Akrap}},
  \bibinfo {author} {\bibfnamefont {H.}~\bibnamefont {Berger}}, \bibinfo
  {author} {\bibfnamefont {L.}~\bibnamefont {Forr{\'o}}},\ and\ \bibinfo
  {author} {\bibfnamefont {E.}~\bibnamefont {Tuti{\v s}}},\ }\bibfield  {title}
  {\bibinfo {title} {From {{Mott}} state to superconductivity in
  {{1$T$-TaS$_2$}}},\ }\href {https://doi.org/10.1038/nmat2318} {\bibfield
  {journal} {\bibinfo  {journal} {Nat. Mater.}\ }\textbf {\bibinfo {volume}
  {7}},\ \bibinfo {pages} {960} (\bibinfo {year} {2008})}\BibitemShut {NoStop}%
\bibitem [{\citenamefont
  {Rossnagel}(2011)}]{rossnagelOriginChargedensityWaves2011}%
  \BibitemOpen
  \bibfield  {author} {\bibinfo {author} {\bibfnamefont {K.}~\bibnamefont
  {Rossnagel}},\ }\bibfield  {title} {\bibinfo {title} {On the origin of
  charge-density waves in select layered transition-metal dichalcogenides},\
  }\href {https://doi.org/10.1088/0953-8984/23/21/213001} {\bibfield  {journal}
  {\bibinfo  {journal} {J. Phys. Condens. Matter}\ }\textbf {\bibinfo {volume}
  {23}},\ \bibinfo {pages} {213001} (\bibinfo {year} {2011})}\BibitemShut
  {NoStop}%
\bibitem [{\citenamefont {Wang}\ \emph {et~al.}(2020)\citenamefont {Wang},
  \citenamefont {Yao}, \citenamefont {Xin}, \citenamefont {Han}, \citenamefont
  {Wang}, \citenamefont {Chen}, \citenamefont {Cai}, \citenamefont {Li},\ and\
  \citenamefont {Zhang}}]{wangBandInsulatorMott2020}%
  \BibitemOpen
  \bibfield  {author} {\bibinfo {author} {\bibfnamefont {Y.~D.}\ \bibnamefont
  {Wang}}, \bibinfo {author} {\bibfnamefont {W.~L.}\ \bibnamefont {Yao}},
  \bibinfo {author} {\bibfnamefont {Z.~M.}\ \bibnamefont {Xin}}, \bibinfo
  {author} {\bibfnamefont {T.~T.}\ \bibnamefont {Han}}, \bibinfo {author}
  {\bibfnamefont {Z.~G.}\ \bibnamefont {Wang}}, \bibinfo {author}
  {\bibfnamefont {L.}~\bibnamefont {Chen}}, \bibinfo {author} {\bibfnamefont
  {C.}~\bibnamefont {Cai}}, \bibinfo {author} {\bibfnamefont {Y.}~\bibnamefont
  {Li}},\ and\ \bibinfo {author} {\bibfnamefont {Y.}~\bibnamefont {Zhang}},\
  }\bibfield  {title} {\bibinfo {title} {Band insulator to {{Mott}} insulator
  transition in 1{{{\emph{T}}}}-{{TaS}}{\textsubscript{2}}},\ }\href
  {https://doi.org/10.1038/s41467-020-18040-4} {\bibfield  {journal} {\bibinfo
  {journal} {Nat. Commun.}\ }\textbf {\bibinfo {volume} {11}},\ \bibinfo
  {pages} {4215} (\bibinfo {year} {2020})}\BibitemShut {NoStop}%
\bibitem [{\citenamefont {Di~Salvo}\ and\ \citenamefont
  {Graebner}(1977)}]{disalvoLowTemperatureElectrical1977}%
  \BibitemOpen
  \bibfield  {author} {\bibinfo {author} {\bibfnamefont {F.}~\bibnamefont
  {Di~Salvo}}\ and\ \bibinfo {author} {\bibfnamefont {J.}~\bibnamefont
  {Graebner}},\ }\bibfield  {title} {\bibinfo {title} {The low temperature
  electrical properties of {{1$T$-TaS$_2$}}},\ }\href
  {https://doi.org/10.1016/0038-1098(77)90961-9} {\bibfield  {journal}
  {\bibinfo  {journal} {Solid State Commun.}\ }\textbf {\bibinfo {volume}
  {23}},\ \bibinfo {pages} {825} (\bibinfo {year} {1977})}\BibitemShut
  {NoStop}%
\bibitem [{\citenamefont {Kim}\ \emph {et~al.}(1994)\citenamefont {Kim},
  \citenamefont {Yamaguchi}, \citenamefont {Hasegawa},\ and\ \citenamefont
  {Kitazawa}}]{kimObservationMottLocalization1994}%
  \BibitemOpen
  \bibfield  {author} {\bibinfo {author} {\bibfnamefont {J.-J.}\ \bibnamefont
  {Kim}}, \bibinfo {author} {\bibfnamefont {W.}~\bibnamefont {Yamaguchi}},
  \bibinfo {author} {\bibfnamefont {T.}~\bibnamefont {Hasegawa}},\ and\
  \bibinfo {author} {\bibfnamefont {K.}~\bibnamefont {Kitazawa}},\ }\bibfield
  {title} {\bibinfo {title} {Observation of {{Mott Localization Gap Using Low
  Temperature Scanning Tunneling Spectroscopy}} in {{Commensurate}}
  {{1$T$-TaS$_2$}}},\ }\href {https://doi.org/10.1103/PhysRevLett.73.2103}
  {\bibfield  {journal} {\bibinfo  {journal} {Phys. Rev. Lett.}\ }\textbf
  {\bibinfo {volume} {73}},\ \bibinfo {pages} {2103} (\bibinfo {year}
  {1994})}\BibitemShut {NoStop}%
\bibitem [{\citenamefont {Law}\ and\ \citenamefont
  {Lee}(2017)}]{law1TTaS2Quantum2017}%
  \BibitemOpen
  \bibfield  {author} {\bibinfo {author} {\bibfnamefont {K.~T.}\ \bibnamefont
  {Law}}\ and\ \bibinfo {author} {\bibfnamefont {P.~A.}\ \bibnamefont {Lee}},\
  }\bibfield  {title} {\bibinfo {title}
  {1{{{\emph{T}}}}-{{TaS}}{\textsubscript{2}} as a quantum spin liquid},\
  }\href {https://doi.org/10.1073/pnas.1706769114} {\bibfield  {journal}
  {\bibinfo  {journal} {Proc. Natl. Acad. Sci. U.S.A.}\ }\textbf {\bibinfo
  {volume} {114}},\ \bibinfo {pages} {6996} (\bibinfo {year}
  {2017})}\BibitemShut {NoStop}%
\bibitem [{\citenamefont {Klanj{\v s}ek}\ \emph {et~al.}(2017)\citenamefont
  {Klanj{\v s}ek}, \citenamefont {Zorko}, \citenamefont {{\v Z}itko},
  \citenamefont {Mravlje}, \citenamefont {Jagli{\v c}i{\'c}}, \citenamefont
  {Biswas}, \citenamefont {Prelov{\v s}ek}, \citenamefont {Mihailovic},\ and\
  \citenamefont {Ar{\v c}on}}]{klanjsekHightemperatureQuantumSpin2017}%
  \BibitemOpen
  \bibfield  {author} {\bibinfo {author} {\bibfnamefont {M.}~\bibnamefont
  {Klanj{\v s}ek}}, \bibinfo {author} {\bibfnamefont {A.}~\bibnamefont
  {Zorko}}, \bibinfo {author} {\bibfnamefont {R.}~\bibnamefont {{\v Z}itko}},
  \bibinfo {author} {\bibfnamefont {J.}~\bibnamefont {Mravlje}}, \bibinfo
  {author} {\bibfnamefont {Z.}~\bibnamefont {Jagli{\v c}i{\'c}}}, \bibinfo
  {author} {\bibfnamefont {P.~K.}\ \bibnamefont {Biswas}}, \bibinfo {author}
  {\bibfnamefont {P.}~\bibnamefont {Prelov{\v s}ek}}, \bibinfo {author}
  {\bibfnamefont {D.}~\bibnamefont {Mihailovic}},\ and\ \bibinfo {author}
  {\bibfnamefont {D.}~\bibnamefont {Ar{\v c}on}},\ }\bibfield  {title}
  {\bibinfo {title} {A high-temperature quantum spin liquid with polaron
  spins},\ }\href {https://doi.org/10.1038/nphys4212} {\bibfield  {journal}
  {\bibinfo  {journal} {Nat. Phys.}\ }\textbf {\bibinfo {volume} {13}},\
  \bibinfo {pages} {1130} (\bibinfo {year} {2017})}\BibitemShut {NoStop}%
\bibitem [{\citenamefont {He}\ \emph {et~al.}(2018)\citenamefont {He},
  \citenamefont {Xu}, \citenamefont {Chen}, \citenamefont {Law},\ and\
  \citenamefont {Lee}}]{heSpinonFermiSurface2018}%
  \BibitemOpen
  \bibfield  {author} {\bibinfo {author} {\bibfnamefont {W.-Y.}\ \bibnamefont
  {He}}, \bibinfo {author} {\bibfnamefont {X.~Y.}\ \bibnamefont {Xu}}, \bibinfo
  {author} {\bibfnamefont {G.}~\bibnamefont {Chen}}, \bibinfo {author}
  {\bibfnamefont {K.~T.}\ \bibnamefont {Law}},\ and\ \bibinfo {author}
  {\bibfnamefont {P.~A.}\ \bibnamefont {Lee}},\ }\bibfield  {title} {\bibinfo
  {title} {Spinon {{Fermi Surface}} in a {{Cluster Mott Insulator Model}} on a
  {{Triangular Lattice}} and {{Possible Application}} to {{1$T$-TaS$_2$}}},\
  }\href {https://doi.org/10.1103/PhysRevLett.121.046401} {\bibfield  {journal}
  {\bibinfo  {journal} {Phys. Rev. Lett.}\ }\textbf {\bibinfo {volume} {121}},\
  \bibinfo {pages} {046401} (\bibinfo {year} {2018})}\BibitemShut {NoStop}%
\bibitem [{\citenamefont {Wen}\ \emph {et~al.}(2019)\citenamefont {Wen},
  \citenamefont {Yu}, \citenamefont {Li}, \citenamefont {Yu},\ and\
  \citenamefont {Li}}]{wenExperimentalIdentificationQuantum2019}%
  \BibitemOpen
  \bibfield  {author} {\bibinfo {author} {\bibfnamefont {J.}~\bibnamefont
  {Wen}}, \bibinfo {author} {\bibfnamefont {S.-L.}\ \bibnamefont {Yu}},
  \bibinfo {author} {\bibfnamefont {S.}~\bibnamefont {Li}}, \bibinfo {author}
  {\bibfnamefont {W.}~\bibnamefont {Yu}},\ and\ \bibinfo {author}
  {\bibfnamefont {J.-X.}\ \bibnamefont {Li}},\ }\bibfield  {title} {\bibinfo
  {title} {Experimental identification of quantum spin liquids},\ }\href
  {https://doi.org/10.1038/s41535-019-0151-6} {\bibfield  {journal} {\bibinfo
  {journal} {npj Quantum Mater.}\ }\textbf {\bibinfo {volume} {4}},\ \bibinfo
  {pages} {1} (\bibinfo {year} {2019})}\BibitemShut {NoStop}%
\bibitem [{\citenamefont {Ritschel}\ \emph {et~al.}(2015)\citenamefont
  {Ritschel}, \citenamefont {Trinckauf}, \citenamefont {Koepernik},
  \citenamefont {B{\"u}chner}, \citenamefont {v.~Zimmermann}, \citenamefont
  {Berger}, \citenamefont {Joe}, \citenamefont {Abbamonte},\ and\ \citenamefont
  {Geck}}]{ritschelOrbitalTexturesCharge2015}%
  \BibitemOpen
  \bibfield  {author} {\bibinfo {author} {\bibfnamefont {T.}~\bibnamefont
  {Ritschel}}, \bibinfo {author} {\bibfnamefont {J.}~\bibnamefont {Trinckauf}},
  \bibinfo {author} {\bibfnamefont {K.}~\bibnamefont {Koepernik}}, \bibinfo
  {author} {\bibfnamefont {B.}~\bibnamefont {B{\"u}chner}}, \bibinfo {author}
  {\bibfnamefont {M.}~\bibnamefont {v.~Zimmermann}}, \bibinfo {author}
  {\bibfnamefont {H.}~\bibnamefont {Berger}}, \bibinfo {author} {\bibfnamefont
  {Y.~I.}\ \bibnamefont {Joe}}, \bibinfo {author} {\bibfnamefont
  {P.}~\bibnamefont {Abbamonte}},\ and\ \bibinfo {author} {\bibfnamefont
  {J.}~\bibnamefont {Geck}},\ }\bibfield  {title} {\bibinfo {title} {Orbital
  textures and charge density waves in transition metal dichalcogenides},\
  }\href {https://doi.org/10.1038/nphys3267} {\bibfield  {journal} {\bibinfo
  {journal} {Nat. Phys.}\ }\textbf {\bibinfo {volume} {11}},\ \bibinfo {pages}
  {328} (\bibinfo {year} {2015})}\BibitemShut {NoStop}%
\bibitem [{\citenamefont {Ritschel}\ \emph {et~al.}(2018)\citenamefont
  {Ritschel}, \citenamefont {Berger},\ and\ \citenamefont
  {Geck}}]{ritschelStackingdrivenGapFormation2018}%
  \BibitemOpen
  \bibfield  {author} {\bibinfo {author} {\bibfnamefont {T.}~\bibnamefont
  {Ritschel}}, \bibinfo {author} {\bibfnamefont {H.}~\bibnamefont {Berger}},\
  and\ \bibinfo {author} {\bibfnamefont {J.}~\bibnamefont {Geck}},\ }\bibfield
  {title} {\bibinfo {title} {Stacking-driven gap formation in layered
  {{1$T$-TaS$_2$}}},\ }\href {https://doi.org/10.1103/PhysRevB.98.195134}
  {\bibfield  {journal} {\bibinfo  {journal} {Phys. Rev. B}\ }\textbf {\bibinfo
  {volume} {98}},\ \bibinfo {pages} {195134} (\bibinfo {year}
  {2018})}\BibitemShut {NoStop}%
\bibitem [{\citenamefont {Lee}\ \emph {et~al.}(2019)\citenamefont {Lee},
  \citenamefont {Goh},\ and\ \citenamefont
  {Cho}}]{leeOriginInsulatingPhase2019}%
  \BibitemOpen
  \bibfield  {author} {\bibinfo {author} {\bibfnamefont {S.-H.}\ \bibnamefont
  {Lee}}, \bibinfo {author} {\bibfnamefont {J.~S.}\ \bibnamefont {Goh}},\ and\
  \bibinfo {author} {\bibfnamefont {D.}~\bibnamefont {Cho}},\ }\bibfield
  {title} {\bibinfo {title} {Origin of the {{Insulating Phase}} and
  {{First-Order Metal-Insulator Transition}} in {{1$T$-TaS$_2$}}},\ }\href
  {https://doi.org/10.1103/PhysRevLett.122.106404} {\bibfield  {journal}
  {\bibinfo  {journal} {Phys. Rev. Lett.}\ }\textbf {\bibinfo {volume} {122}},\
  \bibinfo {pages} {106404} (\bibinfo {year} {2019})}\BibitemShut {NoStop}%
\bibitem [{\citenamefont {Butler}\ \emph {et~al.}(2020)\citenamefont {Butler},
  \citenamefont {Yoshida}, \citenamefont {Hanaguri},\ and\ \citenamefont
  {Iwasa}}]{butlerMottnessUnitcellDoubling2020}%
  \BibitemOpen
  \bibfield  {author} {\bibinfo {author} {\bibfnamefont {C.~J.}\ \bibnamefont
  {Butler}}, \bibinfo {author} {\bibfnamefont {M.}~\bibnamefont {Yoshida}},
  \bibinfo {author} {\bibfnamefont {T.}~\bibnamefont {Hanaguri}},\ and\
  \bibinfo {author} {\bibfnamefont {Y.}~\bibnamefont {Iwasa}},\ }\bibfield
  {title} {\bibinfo {title} {Mottness versus unit-cell doubling as the driver
  of the insulating state in {{1$T$-TaS$_2$}}},\ }\href
  {https://doi.org/10.1038/s41467-020-16132-9} {\bibfield  {journal} {\bibinfo
  {journal} {Nat. Commun.}\ }\textbf {\bibinfo {volume} {11}},\ \bibinfo
  {pages} {2477} (\bibinfo {year} {2020})}\BibitemShut {NoStop}%
\bibitem [{\citenamefont {Martino}\ \emph {et~al.}(2020)\citenamefont
  {Martino}, \citenamefont {Pisoni}, \citenamefont {{\'C}iri{\'c}},
  \citenamefont {Arakcheeva}, \citenamefont {Berger}, \citenamefont {Akrap},
  \citenamefont {Putzke}, \citenamefont {Moll}, \citenamefont {Batisti{\'c}},
  \citenamefont {Tuti{\v s}}, \citenamefont {Forr{\'o}},\ and\ \citenamefont
  {Semeniuk}}]{martinoPreferentialOutofplaneConduction2020}%
  \BibitemOpen
  \bibfield  {author} {\bibinfo {author} {\bibfnamefont {E.}~\bibnamefont
  {Martino}}, \bibinfo {author} {\bibfnamefont {A.}~\bibnamefont {Pisoni}},
  \bibinfo {author} {\bibfnamefont {L.}~\bibnamefont {{\'C}iri{\'c}}}, \bibinfo
  {author} {\bibfnamefont {A.}~\bibnamefont {Arakcheeva}}, \bibinfo {author}
  {\bibfnamefont {H.}~\bibnamefont {Berger}}, \bibinfo {author} {\bibfnamefont
  {A.}~\bibnamefont {Akrap}}, \bibinfo {author} {\bibfnamefont
  {C.}~\bibnamefont {Putzke}}, \bibinfo {author} {\bibfnamefont {P.~J.~W.}\
  \bibnamefont {Moll}}, \bibinfo {author} {\bibfnamefont {I.}~\bibnamefont
  {Batisti{\'c}}}, \bibinfo {author} {\bibfnamefont {E.}~\bibnamefont {Tuti{\v
  s}}}, \bibinfo {author} {\bibfnamefont {L.}~\bibnamefont {Forr{\'o}}},\ and\
  \bibinfo {author} {\bibfnamefont {K.}~\bibnamefont {Semeniuk}},\ }\bibfield
  {title} {\bibinfo {title} {Preferential out-of-plane conduction and
  quasi-one-dimensional electronic states in layered {{1$T$-TaS$_2$}}},\ }\href
  {https://doi.org/10.1038/s41699-020-0145-z} {\bibfield  {journal} {\bibinfo
  {journal} {npj 2D Mater. Appl.}\ }\textbf {\bibinfo {volume} {4}},\ \bibinfo
  {pages} {7} (\bibinfo {year} {2020})}\BibitemShut {NoStop}%
\bibitem [{\citenamefont {Lee}\ \emph {et~al.}(2021)\citenamefont {Lee},
  \citenamefont {Jin},\ and\ \citenamefont
  {Yeom}}]{leeDistinguishingMottInsulator2021}%
  \BibitemOpen
  \bibfield  {author} {\bibinfo {author} {\bibfnamefont {J.}~\bibnamefont
  {Lee}}, \bibinfo {author} {\bibfnamefont {K.-H.}\ \bibnamefont {Jin}},\ and\
  \bibinfo {author} {\bibfnamefont {H.~W.}\ \bibnamefont {Yeom}},\ }\bibfield
  {title} {\bibinfo {title} {Distinguishing a {{Mott Insulator}} from a
  {{Trivial Insulator}} with {{Atomic Adsorbates}}},\ }\href
  {https://doi.org/10.1103/PhysRevLett.126.196405} {\bibfield  {journal}
  {\bibinfo  {journal} {Phys. Rev. Lett.}\ }\textbf {\bibinfo {volume} {126}},\
  \bibinfo {pages} {196405} (\bibinfo {year} {2021})}\BibitemShut {NoStop}%
\bibitem [{\citenamefont {Petocchi}\ \emph {et~al.}(2022)\citenamefont
  {Petocchi}, \citenamefont {Nicholson}, \citenamefont {Salzmann},
  \citenamefont {Pasquier}, \citenamefont {Yazyev}, \citenamefont {Monney},\
  and\ \citenamefont {Werner}}]{petocchiMottHybridizationGap2022}%
  \BibitemOpen
  \bibfield  {author} {\bibinfo {author} {\bibfnamefont {F.}~\bibnamefont
  {Petocchi}}, \bibinfo {author} {\bibfnamefont {C.~W.}\ \bibnamefont
  {Nicholson}}, \bibinfo {author} {\bibfnamefont {B.}~\bibnamefont {Salzmann}},
  \bibinfo {author} {\bibfnamefont {D.}~\bibnamefont {Pasquier}}, \bibinfo
  {author} {\bibfnamefont {O.~V.}\ \bibnamefont {Yazyev}}, \bibinfo {author}
  {\bibfnamefont {C.}~\bibnamefont {Monney}},\ and\ \bibinfo {author}
  {\bibfnamefont {P.}~\bibnamefont {Werner}},\ }\bibfield  {title} {\bibinfo
  {title} {Mott versus {{Hybridization Gap}} in the {{Low-Temperature Phase}}
  of 1{{{\emph{T}}}}-{{TaS}}{\textsubscript{2}}},\ }\href
  {https://doi.org/10.1103/PhysRevLett.129.016402} {\bibfield  {journal}
  {\bibinfo  {journal} {Phys. Rev. Lett.}\ }\textbf {\bibinfo {volume} {129}},\
  \bibinfo {pages} {016402} (\bibinfo {year} {2022})}\BibitemShut {NoStop}%
\bibitem [{\citenamefont {Nicholson}\ \emph {et~al.}(2024)\citenamefont
  {Nicholson}, \citenamefont {Petocchi}, \citenamefont {Salzmann},
  \citenamefont {Witteveen}, \citenamefont {Rumo}, \citenamefont {Kremer},
  \citenamefont {Ivashko}, \citenamefont {{von Rohr}}, \citenamefont {Werner},\
  and\ \citenamefont {Monney}}]{nicholsonGapCollapseFlat2024}%
  \BibitemOpen
  \bibfield  {author} {\bibinfo {author} {\bibfnamefont {C.~W.}\ \bibnamefont
  {Nicholson}}, \bibinfo {author} {\bibfnamefont {F.}~\bibnamefont {Petocchi}},
  \bibinfo {author} {\bibfnamefont {B.}~\bibnamefont {Salzmann}}, \bibinfo
  {author} {\bibfnamefont {C.}~\bibnamefont {Witteveen}}, \bibinfo {author}
  {\bibfnamefont {M.}~\bibnamefont {Rumo}}, \bibinfo {author} {\bibfnamefont
  {G.}~\bibnamefont {Kremer}}, \bibinfo {author} {\bibfnamefont
  {O.}~\bibnamefont {Ivashko}}, \bibinfo {author} {\bibfnamefont {F.~O.}\
  \bibnamefont {{von Rohr}}}, \bibinfo {author} {\bibfnamefont
  {P.}~\bibnamefont {Werner}},\ and\ \bibinfo {author} {\bibfnamefont
  {C.}~\bibnamefont {Monney}},\ }\bibfield  {title} {\bibinfo {title} {Gap
  collapse and flat band induced by uniaxial strain in
  1{{{\emph{T}}}}-{{TaS}}{\textsubscript{2}}},\ }\href
  {https://doi.org/10.1103/PhysRevB.109.035167} {\bibfield  {journal} {\bibinfo
   {journal} {Phys. Rev. B}\ }\textbf {\bibinfo {volume} {109}},\ \bibinfo
  {pages} {035167} (\bibinfo {year} {2024})}\BibitemShut {NoStop}%
\bibitem [{\citenamefont {Wang}\ \emph {et~al.}(2024)\citenamefont {Wang},
  \citenamefont {Li}, \citenamefont {Luo}, \citenamefont {Gao}, \citenamefont
  {Han}, \citenamefont {Jiang}, \citenamefont {Tang}, \citenamefont {Ju},
  \citenamefont {Li}, \citenamefont {Lv}, \citenamefont {Cui}, \citenamefont
  {Yang}, \citenamefont {Sun}, \citenamefont {Zhu}, \citenamefont {Gao},
  \citenamefont {Lu}, \citenamefont {Sun}, \citenamefont {Xu}, \citenamefont
  {Xiong},\ and\ \citenamefont {Cao}}]{wangDualisticInsulatorStates2024}%
  \BibitemOpen
  \bibfield  {author} {\bibinfo {author} {\bibfnamefont {Y.}~\bibnamefont
  {Wang}}, \bibinfo {author} {\bibfnamefont {Z.}~\bibnamefont {Li}}, \bibinfo
  {author} {\bibfnamefont {X.}~\bibnamefont {Luo}}, \bibinfo {author}
  {\bibfnamefont {J.}~\bibnamefont {Gao}}, \bibinfo {author} {\bibfnamefont
  {Y.}~\bibnamefont {Han}}, \bibinfo {author} {\bibfnamefont {J.}~\bibnamefont
  {Jiang}}, \bibinfo {author} {\bibfnamefont {J.}~\bibnamefont {Tang}},
  \bibinfo {author} {\bibfnamefont {H.}~\bibnamefont {Ju}}, \bibinfo {author}
  {\bibfnamefont {T.}~\bibnamefont {Li}}, \bibinfo {author} {\bibfnamefont
  {R.}~\bibnamefont {Lv}}, \bibinfo {author} {\bibfnamefont {S.}~\bibnamefont
  {Cui}}, \bibinfo {author} {\bibfnamefont {Y.}~\bibnamefont {Yang}}, \bibinfo
  {author} {\bibfnamefont {Y.}~\bibnamefont {Sun}}, \bibinfo {author}
  {\bibfnamefont {J.}~\bibnamefont {Zhu}}, \bibinfo {author} {\bibfnamefont
  {X.}~\bibnamefont {Gao}}, \bibinfo {author} {\bibfnamefont {W.}~\bibnamefont
  {Lu}}, \bibinfo {author} {\bibfnamefont {Z.}~\bibnamefont {Sun}}, \bibinfo
  {author} {\bibfnamefont {H.}~\bibnamefont {Xu}}, \bibinfo {author}
  {\bibfnamefont {Y.}~\bibnamefont {Xiong}},\ and\ \bibinfo {author}
  {\bibfnamefont {L.}~\bibnamefont {Cao}},\ }\bibfield  {title} {\bibinfo
  {title} {Dualistic insulator states in
  1{{{\emph{T}}}}-{{TaS}}{\textsubscript{2}} crystals},\ }\href
  {https://doi.org/10.1038/s41467-024-47728-0} {\bibfield  {journal} {\bibinfo
  {journal} {Nat. Commun.}\ }\textbf {\bibinfo {volume} {15}},\ \bibinfo
  {pages} {3425} (\bibinfo {year} {2024})}\BibitemShut {NoStop}%
\bibitem [{\citenamefont {Stahl}\ \emph {et~al.}(2020)\citenamefont {Stahl},
  \citenamefont {Kusch}, \citenamefont {Heinsch}, \citenamefont {Garbarino},
  \citenamefont {Kretzschmar}, \citenamefont {Hanff}, \citenamefont
  {Rossnagel}, \citenamefont {Geck},\ and\ \citenamefont
  {Ritschel}}]{stahlCollapseLayerDimerization2020}%
  \BibitemOpen
  \bibfield  {author} {\bibinfo {author} {\bibfnamefont {Q.}~\bibnamefont
  {Stahl}}, \bibinfo {author} {\bibfnamefont {M.}~\bibnamefont {Kusch}},
  \bibinfo {author} {\bibfnamefont {F.}~\bibnamefont {Heinsch}}, \bibinfo
  {author} {\bibfnamefont {G.}~\bibnamefont {Garbarino}}, \bibinfo {author}
  {\bibfnamefont {N.}~\bibnamefont {Kretzschmar}}, \bibinfo {author}
  {\bibfnamefont {K.}~\bibnamefont {Hanff}}, \bibinfo {author} {\bibfnamefont
  {K.}~\bibnamefont {Rossnagel}}, \bibinfo {author} {\bibfnamefont
  {J.}~\bibnamefont {Geck}},\ and\ \bibinfo {author} {\bibfnamefont
  {T.}~\bibnamefont {Ritschel}},\ }\bibfield  {title} {\bibinfo {title}
  {Collapse of layer dimerization in the photo-induced hidden state of
  {{1$T$-TaS$_2$}}},\ }\href {https://doi.org/10.1038/s41467-020-15079-1}
  {\bibfield  {journal} {\bibinfo  {journal} {Nat. Commun.}\ }\textbf {\bibinfo
  {volume} {11}},\ \bibinfo {pages} {1247} (\bibinfo {year}
  {2020})}\BibitemShut {NoStop}%
\bibitem [{\citenamefont {Ramos}\ \emph {et~al.}(2023)\citenamefont {Ramos},
  \citenamefont {Carvalho}, \citenamefont {Monteiro~Lobato}, \citenamefont
  {{Ribeiro-Soares}}, \citenamefont {Fantini}, \citenamefont {Ribeiro},
  \citenamefont {Molino}, \citenamefont {Plumadore}, \citenamefont {Heinz},
  \citenamefont {{Luican-Mayer}},\ and\ \citenamefont
  {Pimenta}}]{ramosSelectiveElectronPhonon2023}%
  \BibitemOpen
  \bibfield  {author} {\bibinfo {author} {\bibfnamefont {S.~L. L.~M.}\
  \bibnamefont {Ramos}}, \bibinfo {author} {\bibfnamefont {B.~R.}\ \bibnamefont
  {Carvalho}}, \bibinfo {author} {\bibfnamefont {R.~L.}\ \bibnamefont
  {Monteiro~Lobato}}, \bibinfo {author} {\bibfnamefont {J.}~\bibnamefont
  {{Ribeiro-Soares}}}, \bibinfo {author} {\bibfnamefont {C.}~\bibnamefont
  {Fantini}}, \bibinfo {author} {\bibfnamefont {H.~B.}\ \bibnamefont
  {Ribeiro}}, \bibinfo {author} {\bibfnamefont {L.}~\bibnamefont {Molino}},
  \bibinfo {author} {\bibfnamefont {R.}~\bibnamefont {Plumadore}}, \bibinfo
  {author} {\bibfnamefont {T.}~\bibnamefont {Heinz}}, \bibinfo {author}
  {\bibfnamefont {A.}~\bibnamefont {{Luican-Mayer}}},\ and\ \bibinfo {author}
  {\bibfnamefont {M.~A.}\ \bibnamefont {Pimenta}},\ }\bibfield  {title}
  {\bibinfo {title} {Selective {{Electron}}--{{Phonon Coupling}} in
  {{Dimerized}} 1{{{\emph{T}}}}-{{TaS}}{\textsubscript{2}} {{Revealed}} by
  {{Resonance Raman Spectroscopy}}},\ }\href
  {https://doi.org/10.1021/acsnano.3c03902} {\bibfield  {journal} {\bibinfo
  {journal} {ACS Nano}\ }\textbf {\bibinfo {volume} {17}},\ \bibinfo {pages}
  {15883} (\bibinfo {year} {2023})}\BibitemShut {NoStop}%
\bibitem [{\citenamefont {Ma}\ \emph {et~al.}(2016)\citenamefont {Ma},
  \citenamefont {Ye}, \citenamefont {Yu}, \citenamefont {Lu}, \citenamefont
  {Niu}, \citenamefont {Kim}, \citenamefont {Feng}, \citenamefont
  {Tom{\'a}nek}, \citenamefont {Son}, \citenamefont {Chen},\ and\ \citenamefont
  {Zhang}}]{maMetallicMosaicPhase2016}%
  \BibitemOpen
  \bibfield  {author} {\bibinfo {author} {\bibfnamefont {L.}~\bibnamefont
  {Ma}}, \bibinfo {author} {\bibfnamefont {C.}~\bibnamefont {Ye}}, \bibinfo
  {author} {\bibfnamefont {Y.}~\bibnamefont {Yu}}, \bibinfo {author}
  {\bibfnamefont {X.~F.}\ \bibnamefont {Lu}}, \bibinfo {author} {\bibfnamefont
  {X.}~\bibnamefont {Niu}}, \bibinfo {author} {\bibfnamefont {S.}~\bibnamefont
  {Kim}}, \bibinfo {author} {\bibfnamefont {D.}~\bibnamefont {Feng}}, \bibinfo
  {author} {\bibfnamefont {D.}~\bibnamefont {Tom{\'a}nek}}, \bibinfo {author}
  {\bibfnamefont {Y.-W.}\ \bibnamefont {Son}}, \bibinfo {author} {\bibfnamefont
  {X.~H.}\ \bibnamefont {Chen}},\ and\ \bibinfo {author} {\bibfnamefont
  {Y.}~\bibnamefont {Zhang}},\ }\bibfield  {title} {\bibinfo {title} {A
  metallic mosaic phase and the origin of {{Mott-insulating}} state in
  {{1$T$-TaS$_2$}}},\ }\href {https://doi.org/10.1038/ncomms10956} {\bibfield
  {journal} {\bibinfo  {journal} {Nat. Commun.}\ }\textbf {\bibinfo {volume}
  {7}},\ \bibinfo {pages} {10956} (\bibinfo {year} {2016})}\BibitemShut
  {NoStop}%
\bibitem [{\citenamefont {Wu}\ \emph {et~al.}(2022)\citenamefont {Wu},
  \citenamefont {Bu}, \citenamefont {Zhang}, \citenamefont {Fei}, \citenamefont
  {Zheng}, \citenamefont {Gao}, \citenamefont {Luo}, \citenamefont {Liu},
  \citenamefont {Sun},\ and\ \citenamefont {Yin}}]{wuEffectStackingOrder2022}%
  \BibitemOpen
  \bibfield  {author} {\bibinfo {author} {\bibfnamefont {Z.}~\bibnamefont
  {Wu}}, \bibinfo {author} {\bibfnamefont {K.}~\bibnamefont {Bu}}, \bibinfo
  {author} {\bibfnamefont {W.}~\bibnamefont {Zhang}}, \bibinfo {author}
  {\bibfnamefont {Y.}~\bibnamefont {Fei}}, \bibinfo {author} {\bibfnamefont
  {Y.}~\bibnamefont {Zheng}}, \bibinfo {author} {\bibfnamefont
  {J.}~\bibnamefont {Gao}}, \bibinfo {author} {\bibfnamefont {X.}~\bibnamefont
  {Luo}}, \bibinfo {author} {\bibfnamefont {Z.}~\bibnamefont {Liu}}, \bibinfo
  {author} {\bibfnamefont {Y.-P.}\ \bibnamefont {Sun}},\ and\ \bibinfo {author}
  {\bibfnamefont {Y.}~\bibnamefont {Yin}},\ }\bibfield  {title} {\bibinfo
  {title} {Effect of stacking order on the electronic state of
  1{{{\emph{T}}}}-{{TaS}}{\textsubscript{2}}},\ }\href
  {https://doi.org/10.1103/PhysRevB.105.035109} {\bibfield  {journal} {\bibinfo
   {journal} {Phys. Rev. B}\ }\textbf {\bibinfo {volume} {105}},\ \bibinfo
  {pages} {035109} (\bibinfo {year} {2022})}\BibitemShut {NoStop}%
\bibitem [{\citenamefont {Zhang}\ \emph {et~al.}(2022)\citenamefont {Zhang},
  \citenamefont {Wu}, \citenamefont {Bu}, \citenamefont {Fei}, \citenamefont
  {Zheng}, \citenamefont {Gao}, \citenamefont {Luo}, \citenamefont {Liu},
  \citenamefont {Sun},\ and\ \citenamefont
  {Yin}}]{zhangReconcilingBulkMetallic2022}%
  \BibitemOpen
  \bibfield  {author} {\bibinfo {author} {\bibfnamefont {W.}~\bibnamefont
  {Zhang}}, \bibinfo {author} {\bibfnamefont {Z.}~\bibnamefont {Wu}}, \bibinfo
  {author} {\bibfnamefont {K.}~\bibnamefont {Bu}}, \bibinfo {author}
  {\bibfnamefont {Y.}~\bibnamefont {Fei}}, \bibinfo {author} {\bibfnamefont
  {Y.}~\bibnamefont {Zheng}}, \bibinfo {author} {\bibfnamefont
  {J.}~\bibnamefont {Gao}}, \bibinfo {author} {\bibfnamefont {X.}~\bibnamefont
  {Luo}}, \bibinfo {author} {\bibfnamefont {Z.}~\bibnamefont {Liu}}, \bibinfo
  {author} {\bibfnamefont {Y.-P.}\ \bibnamefont {Sun}},\ and\ \bibinfo {author}
  {\bibfnamefont {Y.}~\bibnamefont {Yin}},\ }\bibfield  {title} {\bibinfo
  {title} {Reconciling the bulk metallic and surface insulating state in
  1{{{\emph{T}}}}-{{TaS}}{\textsubscript{2}}},\ }\href
  {https://doi.org/10.1103/PhysRevB.105.035110} {\bibfield  {journal} {\bibinfo
   {journal} {Phys. Rev. B}\ }\textbf {\bibinfo {volume} {105}},\ \bibinfo
  {pages} {035110} (\bibinfo {year} {2022})}\BibitemShut {NoStop}%
\bibitem [{\citenamefont {Lee}\ and\ \citenamefont
  {Cho}(2023)}]{leeChargeDensityWave2023}%
  \BibitemOpen
  \bibfield  {author} {\bibinfo {author} {\bibfnamefont {S.-H.}\ \bibnamefont
  {Lee}}\ and\ \bibinfo {author} {\bibfnamefont {D.}~\bibnamefont {Cho}},\
  }\bibfield  {title} {\bibinfo {title} {Charge density wave surface
  reconstruction in a van der {{Waals}} layered material},\ }\href
  {https://doi.org/10.1038/s41467-023-41500-6} {\bibfield  {journal} {\bibinfo
  {journal} {Nat. Commun.}\ }\textbf {\bibinfo {volume} {14}},\ \bibinfo
  {pages} {5735} (\bibinfo {year} {2023})}\BibitemShut {NoStop}%
\bibitem [{\citenamefont {Liao}(2006)}]{liaoPracticalElectronMicroscopy2006}%
  \BibitemOpen
  \bibfield  {author} {\bibinfo {author} {\bibfnamefont {Y.}~\bibnamefont
  {Liao}},\ }\href@noop {} {\emph {\bibinfo {title} {Practical {{Electron
  Microscopy}} and {{Database}}}}}\ (\bibinfo  {publisher} {GlobalSino},\
  \bibinfo {year} {2006})\BibitemShut {NoStop}%
\bibitem [{\citenamefont {Politano}\ \emph {et~al.}(2018)\citenamefont
  {Politano}, \citenamefont {Chiarello}, \citenamefont {Ghosh}, \citenamefont
  {Sadhukhan}, \citenamefont {Kuo}, \citenamefont {Lue}, \citenamefont
  {Pellegrini},\ and\ \citenamefont {Agarwal}}]{politano3DDiracPlasmons2018}%
  \BibitemOpen
  \bibfield  {author} {\bibinfo {author} {\bibfnamefont {A.}~\bibnamefont
  {Politano}}, \bibinfo {author} {\bibfnamefont {G.}~\bibnamefont {Chiarello}},
  \bibinfo {author} {\bibfnamefont {B.}~\bibnamefont {Ghosh}}, \bibinfo
  {author} {\bibfnamefont {K.}~\bibnamefont {Sadhukhan}}, \bibinfo {author}
  {\bibfnamefont {C.-N.}\ \bibnamefont {Kuo}}, \bibinfo {author} {\bibfnamefont
  {C.~S.}\ \bibnamefont {Lue}}, \bibinfo {author} {\bibfnamefont
  {V.}~\bibnamefont {Pellegrini}},\ and\ \bibinfo {author} {\bibfnamefont
  {A.}~\bibnamefont {Agarwal}},\ }\bibfield  {title} {\bibinfo {title} {{{3D
  Dirac Plasmons}} in the {{Type-II Dirac Semimetal
  PtTe}}{\textsubscript{2}}},\ }\href
  {https://doi.org/10.1103/PhysRevLett.121.086804} {\bibfield  {journal}
  {\bibinfo  {journal} {Phys. Rev. Lett.}\ }\textbf {\bibinfo {volume} {121}},\
  \bibinfo {pages} {086804} (\bibinfo {year} {2018})}\BibitemShut {NoStop}%
\bibitem [{\citenamefont {Ghosh}\ \emph {et~al.}(2017)\citenamefont {Ghosh},
  \citenamefont {Kumar}, \citenamefont {Thakur}, \citenamefont {Chauhan},
  \citenamefont {Bhowmick},\ and\ \citenamefont
  {Agarwal}}]{ghoshAnisotropicPlasmonsExcitons2017}%
  \BibitemOpen
  \bibfield  {author} {\bibinfo {author} {\bibfnamefont {B.}~\bibnamefont
  {Ghosh}}, \bibinfo {author} {\bibfnamefont {P.}~\bibnamefont {Kumar}},
  \bibinfo {author} {\bibfnamefont {A.}~\bibnamefont {Thakur}}, \bibinfo
  {author} {\bibfnamefont {Y.~S.}\ \bibnamefont {Chauhan}}, \bibinfo {author}
  {\bibfnamefont {S.}~\bibnamefont {Bhowmick}},\ and\ \bibinfo {author}
  {\bibfnamefont {A.}~\bibnamefont {Agarwal}},\ }\bibfield  {title} {\bibinfo
  {title} {Anisotropic plasmons, excitons, and electron energy loss
  spectroscopy of phosphorene},\ }\href
  {https://doi.org/10.1103/PhysRevB.96.035422} {\bibfield  {journal} {\bibinfo
  {journal} {Phys. Rev. B}\ }\textbf {\bibinfo {volume} {96}},\ \bibinfo
  {pages} {035422} (\bibinfo {year} {2017})}\BibitemShut {NoStop}%
\bibitem [{\citenamefont {Jia}\ \emph {et~al.}(2020)\citenamefont {Jia},
  \citenamefont {Wang}, \citenamefont {Yan}, \citenamefont {Xue}, \citenamefont
  {Zhang}, \citenamefont {Zhou}, \citenamefont {Shi}, \citenamefont {Zhu},
  \citenamefont {Yao},\ and\ \citenamefont
  {Guo}}]{jiaTopologicallyNontrivialInterband2020}%
  \BibitemOpen
  \bibfield  {author} {\bibinfo {author} {\bibfnamefont {X.}~\bibnamefont
  {Jia}}, \bibinfo {author} {\bibfnamefont {M.}~\bibnamefont {Wang}}, \bibinfo
  {author} {\bibfnamefont {D.}~\bibnamefont {Yan}}, \bibinfo {author}
  {\bibfnamefont {S.}~\bibnamefont {Xue}}, \bibinfo {author} {\bibfnamefont
  {S.}~\bibnamefont {Zhang}}, \bibinfo {author} {\bibfnamefont
  {J.}~\bibnamefont {Zhou}}, \bibinfo {author} {\bibfnamefont {Y.}~\bibnamefont
  {Shi}}, \bibinfo {author} {\bibfnamefont {X.}~\bibnamefont {Zhu}}, \bibinfo
  {author} {\bibfnamefont {Y.}~\bibnamefont {Yao}},\ and\ \bibinfo {author}
  {\bibfnamefont {J.}~\bibnamefont {Guo}},\ }\bibfield  {title} {\bibinfo
  {title} {Topologically nontrivial interband plasmons in type-{{II Weyl}}
  semimetal {{MoTe$_2$}}},\ }\href {https://doi.org/10.1088/1367-2630/abbca5}
  {\bibfield  {journal} {\bibinfo  {journal} {New J. Phys.}\ }\textbf {\bibinfo
  {volume} {22}},\ \bibinfo {pages} {103032} (\bibinfo {year}
  {2020})}\BibitemShut {NoStop}%
\bibitem [{\citenamefont {Hesp}\ \emph {et~al.}(2021)\citenamefont {Hesp},
  \citenamefont {Torre}, \citenamefont {{Rodan-Legrain}}, \citenamefont
  {Novelli}, \citenamefont {Cao}, \citenamefont {Carr}, \citenamefont {Fang},
  \citenamefont {Stepanov}, \citenamefont {{Barcons-Ruiz}}, \citenamefont
  {Herzig~Sheinfux}, \citenamefont {Watanabe}, \citenamefont {Taniguchi},
  \citenamefont {Efetov}, \citenamefont {Kaxiras}, \citenamefont
  {{Jarillo-Herrero}}, \citenamefont {Polini},\ and\ \citenamefont
  {Koppens}}]{hespObservationInterbandCollective2021}%
  \BibitemOpen
  \bibfield  {author} {\bibinfo {author} {\bibfnamefont {N.~C.~H.}\
  \bibnamefont {Hesp}}, \bibinfo {author} {\bibfnamefont {I.}~\bibnamefont
  {Torre}}, \bibinfo {author} {\bibfnamefont {D.}~\bibnamefont
  {{Rodan-Legrain}}}, \bibinfo {author} {\bibfnamefont {P.}~\bibnamefont
  {Novelli}}, \bibinfo {author} {\bibfnamefont {Y.}~\bibnamefont {Cao}},
  \bibinfo {author} {\bibfnamefont {S.}~\bibnamefont {Carr}}, \bibinfo {author}
  {\bibfnamefont {S.}~\bibnamefont {Fang}}, \bibinfo {author} {\bibfnamefont
  {P.}~\bibnamefont {Stepanov}}, \bibinfo {author} {\bibfnamefont
  {D.}~\bibnamefont {{Barcons-Ruiz}}}, \bibinfo {author} {\bibfnamefont
  {H.}~\bibnamefont {Herzig~Sheinfux}}, \bibinfo {author} {\bibfnamefont
  {K.}~\bibnamefont {Watanabe}}, \bibinfo {author} {\bibfnamefont
  {T.}~\bibnamefont {Taniguchi}}, \bibinfo {author} {\bibfnamefont {D.~K.}\
  \bibnamefont {Efetov}}, \bibinfo {author} {\bibfnamefont {E.}~\bibnamefont
  {Kaxiras}}, \bibinfo {author} {\bibfnamefont {P.}~\bibnamefont
  {{Jarillo-Herrero}}}, \bibinfo {author} {\bibfnamefont {M.}~\bibnamefont
  {Polini}},\ and\ \bibinfo {author} {\bibfnamefont {F.~H.~L.}\ \bibnamefont
  {Koppens}},\ }\bibfield  {title} {\bibinfo {title} {Observation of interband
  collective excitations in twisted bilayer graphene},\ }\href
  {https://doi.org/10.1038/s41567-021-01327-8} {\bibfield  {journal} {\bibinfo
  {journal} {Nat. Phys.}\ }\textbf {\bibinfo {volume} {17}},\ \bibinfo {pages}
  {1162} (\bibinfo {year} {2021})}\BibitemShut {NoStop}%
\bibitem [{SM()}]{SM}%
  \BibitemOpen
  \href@noop {} {\bibinfo  {journal} {See Supplemental Material at
  http://link.aps.org/. supplemental for details about sample preparation,
  experimental methods, fitting methods, and calculation details, which
  includes
  Refs.\citep{zongUltrafastManipulationMirror2018,heinzExcitationPlasmonsInterband1980,ibachElectronEnergyLoss2013,zhuHighResolutionElectron2015}}\
  }\BibitemShut {NoStop}%
\bibitem [{\citenamefont {Blaha}\ \emph {et~al.}(2020)\citenamefont {Blaha},
  \citenamefont {Schwarz}, \citenamefont {Tran}, \citenamefont {Laskowski},
  \citenamefont {Madsen},\ and\ \citenamefont
  {Marks}}]{blahaWIEN2kAPW+loProgram2020}%
  \BibitemOpen
\bibfield  {journal} {  }\bibfield  {author} {\bibinfo {author} {\bibfnamefont
  {P.}~\bibnamefont {Blaha}}, \bibinfo {author} {\bibfnamefont
  {K.}~\bibnamefont {Schwarz}}, \bibinfo {author} {\bibfnamefont
  {F.}~\bibnamefont {Tran}}, \bibinfo {author} {\bibfnamefont {R.}~\bibnamefont
  {Laskowski}}, \bibinfo {author} {\bibfnamefont {G.~K.~H.}\ \bibnamefont
  {Madsen}},\ and\ \bibinfo {author} {\bibfnamefont {L.~D.}\ \bibnamefont
  {Marks}},\ }\bibfield  {title} {\bibinfo {title} {{{WIEN2k}}: {{An APW}}+lo
  program for calculating the properties of solids},\ }\href
  {https://doi.org/10.1063/1.5143061} {\bibfield  {journal} {\bibinfo
  {journal} {J. Chem. Phys.}\ }\textbf {\bibinfo {volume} {152}},\ \bibinfo
  {pages} {074101} (\bibinfo {year} {2020})}\BibitemShut {NoStop}%
\bibitem [{\citenamefont {Perdew}\ \emph {et~al.}(1996)\citenamefont {Perdew},
  \citenamefont {Burke},\ and\ \citenamefont
  {Ernzerhof}}]{perdewGeneralizedGradientApproximation1996a}%
  \BibitemOpen
  \bibfield  {author} {\bibinfo {author} {\bibfnamefont {J.~P.}\ \bibnamefont
  {Perdew}}, \bibinfo {author} {\bibfnamefont {K.}~\bibnamefont {Burke}},\ and\
  \bibinfo {author} {\bibfnamefont {M.}~\bibnamefont {Ernzerhof}},\ }\bibfield
  {title} {\bibinfo {title} {Generalized {{Gradient Approximation Made
  Simple}}},\ }\href {https://doi.org/10.1103/PhysRevLett.77.3865} {\bibfield
  {journal} {\bibinfo  {journal} {Phy. Rev. Lett.}\ }\textbf {\bibinfo {volume}
  {77}},\ \bibinfo {pages} {3865} (\bibinfo {year} {1996})}\BibitemShut
  {NoStop}%
\bibitem [{\citenamefont {Mostofi}\ \emph {et~al.}(2008)\citenamefont
  {Mostofi}, \citenamefont {Yates}, \citenamefont {Lee}, \citenamefont {Souza},
  \citenamefont {Vanderbilt},\ and\ \citenamefont
  {Marzari}}]{mostofiWannier90ToolObtaining2008}%
  \BibitemOpen
  \bibfield  {author} {\bibinfo {author} {\bibfnamefont {A.~A.}\ \bibnamefont
  {Mostofi}}, \bibinfo {author} {\bibfnamefont {J.~R.}\ \bibnamefont {Yates}},
  \bibinfo {author} {\bibfnamefont {Y.-S.}\ \bibnamefont {Lee}}, \bibinfo
  {author} {\bibfnamefont {I.}~\bibnamefont {Souza}}, \bibinfo {author}
  {\bibfnamefont {D.}~\bibnamefont {Vanderbilt}},\ and\ \bibinfo {author}
  {\bibfnamefont {N.}~\bibnamefont {Marzari}},\ }\bibfield  {title} {\bibinfo
  {title} {Wannier90: {{A}} tool for obtaining maximally-localised {{Wannier}}
  functions},\ }\href {https://doi.org/10.1016/j.cpc.2007.11.016} {\bibfield
  {journal} {\bibinfo  {journal} {Comput. Phys. Commun.}\ }\textbf {\bibinfo
  {volume} {178}},\ \bibinfo {pages} {685} (\bibinfo {year}
  {2008})}\BibitemShut {NoStop}%
\bibitem [{\citenamefont {Seth}\ \emph {et~al.}(2016)\citenamefont {Seth},
  \citenamefont {Krivenko}, \citenamefont {Ferrero},\ and\ \citenamefont
  {Parcollet}}]{sethTRIQSCTHYBContinuoustime2016}%
  \BibitemOpen
  \bibfield  {author} {\bibinfo {author} {\bibfnamefont {P.}~\bibnamefont
  {Seth}}, \bibinfo {author} {\bibfnamefont {I.}~\bibnamefont {Krivenko}},
  \bibinfo {author} {\bibfnamefont {M.}~\bibnamefont {Ferrero}},\ and\ \bibinfo
  {author} {\bibfnamefont {O.}~\bibnamefont {Parcollet}},\ }\bibfield  {title}
  {\bibinfo {title} {{{TRIQS}}/{{CTHYB}}: {{A}} continuous-time quantum {{Monte
  Carlo}} hybridisation expansion solver for quantum impurity problems},\
  }\href {https://doi.org/10.1016/j.cpc.2015.10.023} {\bibfield  {journal}
  {\bibinfo  {journal} {Comput. Phys. Commun.}\ }\textbf {\bibinfo {volume}
  {200}},\ \bibinfo {pages} {274} (\bibinfo {year} {2016})}\BibitemShut
  {NoStop}%
\bibitem [{\citenamefont {Parcollet}\ \emph {et~al.}(2015)\citenamefont
  {Parcollet}, \citenamefont {Ferrero}, \citenamefont {Ayral}, \citenamefont
  {Hafermann}, \citenamefont {Krivenko}, \citenamefont {Messio},\ and\
  \citenamefont {Seth}}]{parcolletTRIQSToolboxResearch2015}%
  \BibitemOpen
  \bibfield  {author} {\bibinfo {author} {\bibfnamefont {O.}~\bibnamefont
  {Parcollet}}, \bibinfo {author} {\bibfnamefont {M.}~\bibnamefont {Ferrero}},
  \bibinfo {author} {\bibfnamefont {T.}~\bibnamefont {Ayral}}, \bibinfo
  {author} {\bibfnamefont {H.}~\bibnamefont {Hafermann}}, \bibinfo {author}
  {\bibfnamefont {I.}~\bibnamefont {Krivenko}}, \bibinfo {author}
  {\bibfnamefont {L.}~\bibnamefont {Messio}},\ and\ \bibinfo {author}
  {\bibfnamefont {P.}~\bibnamefont {Seth}},\ }\bibfield  {title} {\bibinfo
  {title} {{{TRIQS}}: {{A}} toolbox for research on interacting quantum
  systems},\ }\href {https://doi.org/10.1016/j.cpc.2015.04.023} {\bibfield
  {journal} {\bibinfo  {journal} {Comput. Phys. Commun.}\ }\textbf {\bibinfo
  {volume} {196}},\ \bibinfo {pages} {398} (\bibinfo {year}
  {2015})}\BibitemShut {NoStop}%
\bibitem [{\citenamefont {Kraberger}\ \emph {et~al.}(2017)\citenamefont
  {Kraberger}, \citenamefont {Triebl}, \citenamefont {Zingl},\ and\
  \citenamefont {Aichhorn}}]{krabergerMaximumEntropyFormalism2017}%
  \BibitemOpen
  \bibfield  {author} {\bibinfo {author} {\bibfnamefont {G.~J.}\ \bibnamefont
  {Kraberger}}, \bibinfo {author} {\bibfnamefont {R.}~\bibnamefont {Triebl}},
  \bibinfo {author} {\bibfnamefont {M.}~\bibnamefont {Zingl}},\ and\ \bibinfo
  {author} {\bibfnamefont {M.}~\bibnamefont {Aichhorn}},\ }\bibfield  {title}
  {\bibinfo {title} {Maximum entropy formalism for the analytic continuation of
  matrix-valued {{Green}}'s functions},\ }\href
  {https://doi.org/10.1103/PhysRevB.96.155128} {\bibfield  {journal} {\bibinfo
  {journal} {Phys. Rev. B}\ }\textbf {\bibinfo {volume} {96}},\ \bibinfo
  {pages} {155128} (\bibinfo {year} {2017})}\BibitemShut {NoStop}%
\bibitem [{\citenamefont {Dong}\ \emph {et~al.}(2023)\citenamefont {Dong},
  \citenamefont {Qi}, \citenamefont {Shin}, \citenamefont {Cario},
  \citenamefont {Chen}, \citenamefont {Grasset}, \citenamefont {Boschetto},
  \citenamefont {Weis}, \citenamefont {Lample}, \citenamefont {Pastor},
  \citenamefont {Ritschel}, \citenamefont {Marsi}, \citenamefont {Taleb},
  \citenamefont {Park}, \citenamefont {Rubio}, \citenamefont {Papalazarou},\
  and\ \citenamefont {Perfetti}}]{dongDynamicsElectronicStates2023}%
  \BibitemOpen
  \bibfield  {author} {\bibinfo {author} {\bibfnamefont {J.}~\bibnamefont
  {Dong}}, \bibinfo {author} {\bibfnamefont {W.}~\bibnamefont {Qi}}, \bibinfo
  {author} {\bibfnamefont {D.}~\bibnamefont {Shin}}, \bibinfo {author}
  {\bibfnamefont {L.}~\bibnamefont {Cario}}, \bibinfo {author} {\bibfnamefont
  {Z.}~\bibnamefont {Chen}}, \bibinfo {author} {\bibfnamefont {R.}~\bibnamefont
  {Grasset}}, \bibinfo {author} {\bibfnamefont {D.}~\bibnamefont {Boschetto}},
  \bibinfo {author} {\bibfnamefont {M.}~\bibnamefont {Weis}}, \bibinfo {author}
  {\bibfnamefont {P.}~\bibnamefont {Lample}}, \bibinfo {author} {\bibfnamefont
  {E.}~\bibnamefont {Pastor}}, \bibinfo {author} {\bibfnamefont
  {T.}~\bibnamefont {Ritschel}}, \bibinfo {author} {\bibfnamefont
  {M.}~\bibnamefont {Marsi}}, \bibinfo {author} {\bibfnamefont
  {A.}~\bibnamefont {Taleb}}, \bibinfo {author} {\bibfnamefont
  {N.}~\bibnamefont {Park}}, \bibinfo {author} {\bibfnamefont {A.}~\bibnamefont
  {Rubio}}, \bibinfo {author} {\bibfnamefont {E.}~\bibnamefont {Papalazarou}},\
  and\ \bibinfo {author} {\bibfnamefont {L.}~\bibnamefont {Perfetti}},\
  }\bibfield  {title} {\bibinfo {title} {Dynamics of electronic states in the
  insulating intermediate surface phase of
  1{{{\emph{T}}}}-{{TaS}}{\textsubscript{2}}},\ }\href
  {https://doi.org/10.1103/PhysRevB.108.155145} {\bibfield  {journal} {\bibinfo
   {journal} {Phys. Rev. B}\ }\textbf {\bibinfo {volume} {108}},\ \bibinfo
  {pages} {155145} (\bibinfo {year} {2023})}\BibitemShut {NoStop}%
\bibitem [{\citenamefont {Heinz}(1980)}]{heinzExcitationPlasmonsInterband1980}%
  \BibitemOpen
  \bibfield  {author} {\bibinfo {author} {\bibfnamefont {R.}~\bibnamefont
  {Heinz}},\ }\href {https://doi.org/10.1007/BFb0045951} {\emph {\bibinfo
  {title} {Excitation of {{Plasmons}} and {{Interband Transitions}} by
  {{Electrons}}}}},\ \bibinfo {series} {Springer {{Tracts}} in {{Modern
  Physics}} 88}, Vol.~\bibinfo {volume} {88}\ (\bibinfo  {publisher}
  {Springer-Verlag},\ \bibinfo {address} {Berlin/Heidelberg},\ \bibinfo {year}
  {1980})\BibitemShut {NoStop}%
\bibitem [{\citenamefont {Lambin}\ \emph {et~al.}(1985)\citenamefont {Lambin},
  \citenamefont {Vigneron},\ and\ \citenamefont
  {Lucas}}]{lambinElectronenergylossSpectroscopyMultilayered1985}%
  \BibitemOpen
  \bibfield  {author} {\bibinfo {author} {\bibfnamefont {{\relax
  Ph}.}~\bibnamefont {Lambin}}, \bibinfo {author} {\bibfnamefont {J.~P.}\
  \bibnamefont {Vigneron}},\ and\ \bibinfo {author} {\bibfnamefont {A.~A.}\
  \bibnamefont {Lucas}},\ }\bibfield  {title} {\bibinfo {title}
  {Electron-energy-loss spectroscopy of multilayered materials: {{Theoretical}}
  aspects and study of interface optical phonons in semiconductor
  superlattices},\ }\href {https://doi.org/10.1103/PhysRevB.32.8203} {\bibfield
   {journal} {\bibinfo  {journal} {Phys. Rev. B}\ }\textbf {\bibinfo {volume}
  {32}},\ \bibinfo {pages} {8203} (\bibinfo {year} {1985})}\BibitemShut
  {NoStop}%
\bibitem [{\citenamefont {Lambin}\ \emph {et~al.}(1990)\citenamefont {Lambin},
  \citenamefont {Vigneron},\ and\ \citenamefont
  {Lucas}}]{lambinComputationSurfaceElectronenergyloss1990}%
  \BibitemOpen
  \bibfield  {author} {\bibinfo {author} {\bibfnamefont {{\relax
  Ph}.}~\bibnamefont {Lambin}}, \bibinfo {author} {\bibfnamefont {J.~P.}\
  \bibnamefont {Vigneron}},\ and\ \bibinfo {author} {\bibfnamefont {A.~A.}\
  \bibnamefont {Lucas}},\ }\bibfield  {title} {\bibinfo {title} {Computation of
  the surface electron-energy-loss spectrum in specular geometry for an
  arbitrary plane-stratified medium},\ }\href
  {https://doi.org/10.1016/0010-4655(90)90034-X} {\bibfield  {journal}
  {\bibinfo  {journal} {Comput. Phys. Commun.}\ }\textbf {\bibinfo {volume}
  {60}},\ \bibinfo {pages} {351} (\bibinfo {year} {1990})}\BibitemShut
  {NoStop}%
\bibitem [{\citenamefont {Ibach}\ and\ \citenamefont
  {Mills}(2013)}]{ibachElectronEnergyLoss2013}%
  \BibitemOpen
  \bibfield  {author} {\bibinfo {author} {\bibfnamefont {H.}~\bibnamefont
  {Ibach}}\ and\ \bibinfo {author} {\bibfnamefont {D.~L.}\ \bibnamefont
  {Mills}},\ }\href@noop {} {\emph {\bibinfo {title} {Electron {{Energy Loss
  Spectroscopy}} and {{Surface Vibrations}}}}}\ (\bibinfo  {publisher}
  {Academic Press},\ \bibinfo {address} {New York},\ \bibinfo {year}
  {2013})\BibitemShut {NoStop}%
\bibitem [{\citenamefont {Li}\ \emph {et~al.}(2022)\citenamefont {Li},
  \citenamefont {Lin}, \citenamefont {Miao}, \citenamefont {Zhong},
  \citenamefont {Xue}, \citenamefont {Li}, \citenamefont {Tao}, \citenamefont
  {Wang}, \citenamefont {Guo},\ and\ \citenamefont
  {Zhu}}]{liGeometricEffectHighresolution2022a}%
  \BibitemOpen
  \bibfield  {author} {\bibinfo {author} {\bibfnamefont {J.}~\bibnamefont
  {Li}}, \bibinfo {author} {\bibfnamefont {Z.}~\bibnamefont {Lin}}, \bibinfo
  {author} {\bibfnamefont {G.}~\bibnamefont {Miao}}, \bibinfo {author}
  {\bibfnamefont {W.}~\bibnamefont {Zhong}}, \bibinfo {author} {\bibfnamefont
  {S.}~\bibnamefont {Xue}}, \bibinfo {author} {\bibfnamefont {Y.}~\bibnamefont
  {Li}}, \bibinfo {author} {\bibfnamefont {Z.}~\bibnamefont {Tao}}, \bibinfo
  {author} {\bibfnamefont {W.}~\bibnamefont {Wang}}, \bibinfo {author}
  {\bibfnamefont {J.}~\bibnamefont {Guo}},\ and\ \bibinfo {author}
  {\bibfnamefont {X.}~\bibnamefont {Zhu}},\ }\bibfield  {title} {\bibinfo
  {title} {Geometric effect of high-resolution electron energy loss
  spectroscopy on the identification of plasmons: {{An}} example of graphene},\
  }\href {https://doi.org/10.1016/j.susc.2022.122067} {\bibfield  {journal}
  {\bibinfo  {journal} {Surf. Sci.}\ }\textbf {\bibinfo {volume} {721}},\
  \bibinfo {pages} {122067} (\bibinfo {year} {2022})}\BibitemShut {NoStop}%
\bibitem [{\citenamefont {Lucovsky}\ \emph {et~al.}(1976)\citenamefont
  {Lucovsky}, \citenamefont {Liang}, \citenamefont {White},\ and\ \citenamefont
  {Pisharody}}]{lucovskyReflectivityStudiesTi1976}%
  \BibitemOpen
  \bibfield  {author} {\bibinfo {author} {\bibfnamefont {G.}~\bibnamefont
  {Lucovsky}}, \bibinfo {author} {\bibfnamefont {W.}~\bibnamefont {Liang}},
  \bibinfo {author} {\bibfnamefont {R.}~\bibnamefont {White}},\ and\ \bibinfo
  {author} {\bibfnamefont {K.}~\bibnamefont {Pisharody}},\ }\bibfield  {title}
  {\bibinfo {title} {Reflectivity studies of {{Ti-}} and
  {{Ta-dichalcogenides}}: {{Phonons}}},\ }\href
  {https://doi.org/10.1016/0038-1098(76)91337-5} {\bibfield  {journal}
  {\bibinfo  {journal} {Solid State Commun.}\ }\textbf {\bibinfo {volume}
  {19}},\ \bibinfo {pages} {303} (\bibinfo {year} {1976})}\BibitemShut
  {NoStop}%
\bibitem [{\citenamefont {Dean}\ \emph {et~al.}(2011)\citenamefont {Dean},
  \citenamefont {Petersen}, \citenamefont {Fausti}, \citenamefont {Tobey},
  \citenamefont {Kaiser}, \citenamefont {Gasparov}, \citenamefont {Berger},\
  and\ \citenamefont {Cavalleri}}]{deanPolaronicConductivityPhotoinduced2011}%
  \BibitemOpen
  \bibfield  {author} {\bibinfo {author} {\bibfnamefont {N.}~\bibnamefont
  {Dean}}, \bibinfo {author} {\bibfnamefont {J.~C.}\ \bibnamefont {Petersen}},
  \bibinfo {author} {\bibfnamefont {D.}~\bibnamefont {Fausti}}, \bibinfo
  {author} {\bibfnamefont {R.~I.}\ \bibnamefont {Tobey}}, \bibinfo {author}
  {\bibfnamefont {S.}~\bibnamefont {Kaiser}}, \bibinfo {author} {\bibfnamefont
  {L.~V.}\ \bibnamefont {Gasparov}}, \bibinfo {author} {\bibfnamefont
  {H.}~\bibnamefont {Berger}},\ and\ \bibinfo {author} {\bibfnamefont
  {A.}~\bibnamefont {Cavalleri}},\ }\bibfield  {title} {\bibinfo {title}
  {Polaronic {{Conductivity}} in the {{Photoinduced Phase}} of
  {{1$T$-TaS$_2$}}},\ }\href {https://doi.org/10.1103/PhysRevLett.106.016401}
  {\bibfield  {journal} {\bibinfo  {journal} {Phys. Rev. Lett.}\ }\textbf
  {\bibinfo {volume} {106}},\ \bibinfo {pages} {016401} (\bibinfo {year}
  {2011})}\BibitemShut {NoStop}%
\bibitem [{\citenamefont
  {Velebit}(2015)}]{velebitEffectsSuperstructuringOptical2015}%
  \BibitemOpen
  \bibfield  {author} {\bibinfo {author} {\bibfnamefont {K.}~\bibnamefont
  {Velebit}},\ }\emph {\bibinfo {title} {Effects of Superstructuring on Optical
  and Transport Properties of Selected Layered Materials}},\ \href
  {https://urn.nsk.hr/urn:nbn:hr:217:613872} {Ph.D. thesis},\ \bibinfo
  {school} {University of Zagreb. Faculty of Science. Department of Physics}
  (\bibinfo {year} {2015})\BibitemShut {NoStop}%
\bibitem [{\citenamefont {Tanner}(2015)}]{tannerUseXrayScattering2015}%
  \BibitemOpen
  \bibfield  {author} {\bibinfo {author} {\bibfnamefont {D.~B.}\ \bibnamefont
  {Tanner}},\ }\bibfield  {title} {\bibinfo {title} {Use of {{X}}-ray
  scattering functions in {{Kramers-Kronig}} analysis of reflectance},\ }\href
  {https://doi.org/10.1103/PhysRevB.91.035123} {\bibfield  {journal} {\bibinfo
  {journal} {Phys. Rev. B}\ }\textbf {\bibinfo {volume} {91}},\ \bibinfo
  {pages} {035123} (\bibinfo {year} {2015})}\BibitemShut {NoStop}%
\bibitem [{Note1()}]{Note1}%
  \BibitemOpen
  \bibinfo {note} {Collective excitations related to atomic positions, such as
  phonons and phase modes of charge density waves, can be measured at low
  energies. However, common collective modes within the energy range we
  measured, namely plasmons, are not reflected in the optical
  conductivity.}\BibitemShut {Stop}%
\bibitem [{\citenamefont {Yu}\ \emph {et~al.}(2017)\citenamefont {Yu},
  \citenamefont {Liu}, \citenamefont {Quan}, \citenamefont {Wu}, \citenamefont
  {Lin}, \citenamefont {Chang},\ and\ \citenamefont
  {Zou}}]{yuElectronicCorrelationEffects2017}%
  \BibitemOpen
  \bibfield  {author} {\bibinfo {author} {\bibfnamefont {X.-L.}\ \bibnamefont
  {Yu}}, \bibinfo {author} {\bibfnamefont {D.-Y.}\ \bibnamefont {Liu}},
  \bibinfo {author} {\bibfnamefont {Y.-M.}\ \bibnamefont {Quan}}, \bibinfo
  {author} {\bibfnamefont {J.}~\bibnamefont {Wu}}, \bibinfo {author}
  {\bibfnamefont {H.-Q.}\ \bibnamefont {Lin}}, \bibinfo {author} {\bibfnamefont
  {K.}~\bibnamefont {Chang}},\ and\ \bibinfo {author} {\bibfnamefont {L.-J.}\
  \bibnamefont {Zou}},\ }\bibfield  {title} {\bibinfo {title} {Electronic
  correlation effects and orbital density wave in the layered compound
  {{1$T$-TaS$_2$}}},\ }\href {https://doi.org/10.1103/PhysRevB.96.125138}
  {\bibfield  {journal} {\bibinfo  {journal} {Phys. Rev. B}\ }\textbf {\bibinfo
  {volume} {96}},\ \bibinfo {pages} {125138} (\bibinfo {year}
  {2017})}\BibitemShut {NoStop}%
\bibitem [{\citenamefont {Tian}\ \emph {et~al.}(2024)\citenamefont {Tian},
  \citenamefont {Ding}, \citenamefont {Qiu}, \citenamefont {Meng},
  \citenamefont {Wang}, \citenamefont {Yu}, \citenamefont {Mu}, \citenamefont
  {Wang}, \citenamefont {Sun},\ and\ \citenamefont
  {Zhang}}]{tianElectronicStructuresMott2024}%
  \BibitemOpen
  \bibfield  {author} {\bibinfo {author} {\bibfnamefont {Q.}~\bibnamefont
  {Tian}}, \bibinfo {author} {\bibfnamefont {C.}~\bibnamefont {Ding}}, \bibinfo
  {author} {\bibfnamefont {X.}~\bibnamefont {Qiu}}, \bibinfo {author}
  {\bibfnamefont {Q.}~\bibnamefont {Meng}}, \bibinfo {author} {\bibfnamefont
  {K.}~\bibnamefont {Wang}}, \bibinfo {author} {\bibfnamefont {F.}~\bibnamefont
  {Yu}}, \bibinfo {author} {\bibfnamefont {Y.}~\bibnamefont {Mu}}, \bibinfo
  {author} {\bibfnamefont {C.}~\bibnamefont {Wang}}, \bibinfo {author}
  {\bibfnamefont {J.}~\bibnamefont {Sun}},\ and\ \bibinfo {author}
  {\bibfnamefont {Y.}~\bibnamefont {Zhang}},\ }\bibfield  {title} {\bibinfo
  {title} {Electronic structures and {{Mott}} state of epitaxial
  {{TaS}}{\textsubscript{2}} monolayers},\ }\href
  {https://doi.org/10.1007/s11433-023-2328-1} {\bibfield  {journal} {\bibinfo
  {journal} {Sci. China Phys. Mech. Astron.}\ }\textbf {\bibinfo {volume}
  {67}},\ \bibinfo {pages} {256811} (\bibinfo {year} {2024})}\BibitemShut
  {NoStop}%
\bibitem [{\citenamefont {Stojchevska}\ \emph {et~al.}(2014)\citenamefont
  {Stojchevska}, \citenamefont {Vaskivskyi}, \citenamefont {Mertelj},
  \citenamefont {Kusar}, \citenamefont {Svetin}, \citenamefont {Brazovskii},\
  and\ \citenamefont {Mihailovic}}]{stojchevskaUltrafastSwitchingStable2014}%
  \BibitemOpen
  \bibfield  {author} {\bibinfo {author} {\bibfnamefont {L.}~\bibnamefont
  {Stojchevska}}, \bibinfo {author} {\bibfnamefont {I.}~\bibnamefont
  {Vaskivskyi}}, \bibinfo {author} {\bibfnamefont {T.}~\bibnamefont {Mertelj}},
  \bibinfo {author} {\bibfnamefont {P.}~\bibnamefont {Kusar}}, \bibinfo
  {author} {\bibfnamefont {D.}~\bibnamefont {Svetin}}, \bibinfo {author}
  {\bibfnamefont {S.}~\bibnamefont {Brazovskii}},\ and\ \bibinfo {author}
  {\bibfnamefont {D.}~\bibnamefont {Mihailovic}},\ }\bibfield  {title}
  {\bibinfo {title} {Ultrafast {{Switching}} to a {{Stable Hidden Quantum
  State}} in an {{Electronic Crystal}}},\ }\href
  {https://doi.org/10.1126/science.1241591} {\bibfield  {journal} {\bibinfo
  {journal} {Science}\ }\textbf {\bibinfo {volume} {344}},\ \bibinfo {pages}
  {177} (\bibinfo {year} {2014})}\BibitemShut {NoStop}%
\bibitem [{\citenamefont {Gerasimenko}\ \emph {et~al.}(2019)\citenamefont
  {Gerasimenko}, \citenamefont {Karpov}, \citenamefont {Vaskivskyi},
  \citenamefont {Brazovskii},\ and\ \citenamefont
  {Mihailovic}}]{gerasimenkoIntertwinedChiralCharge2019}%
  \BibitemOpen
  \bibfield  {author} {\bibinfo {author} {\bibfnamefont {Y.~A.}\ \bibnamefont
  {Gerasimenko}}, \bibinfo {author} {\bibfnamefont {P.}~\bibnamefont {Karpov}},
  \bibinfo {author} {\bibfnamefont {I.}~\bibnamefont {Vaskivskyi}}, \bibinfo
  {author} {\bibfnamefont {S.}~\bibnamefont {Brazovskii}},\ and\ \bibinfo
  {author} {\bibfnamefont {D.}~\bibnamefont {Mihailovic}},\ }\bibfield  {title}
  {\bibinfo {title} {Intertwined chiral charge orders and topological
  stabilization of the light-induced state of a prototypical transition metal
  dichalcogenide},\ }\href {https://doi.org/10.1038/s41535-019-0172-1}
  {\bibfield  {journal} {\bibinfo  {journal} {npj Quantum Mater.}\ }\textbf
  {\bibinfo {volume} {4}},\ \bibinfo {pages} {1} (\bibinfo {year}
  {2019})}\BibitemShut {NoStop}%
\bibitem [{\citenamefont {Ma}\ \emph {et~al.}(2019)\citenamefont {Ma},
  \citenamefont {Wang}, \citenamefont {Hou}, \citenamefont {Wu}, \citenamefont
  {Lu},\ and\ \citenamefont {Petrovic}}]{maObservationMultipleMetastable2019}%
  \BibitemOpen
  \bibfield  {author} {\bibinfo {author} {\bibfnamefont {Y.}~\bibnamefont
  {Ma}}, \bibinfo {author} {\bibfnamefont {Z.}~\bibnamefont {Wang}}, \bibinfo
  {author} {\bibfnamefont {Y.}~\bibnamefont {Hou}}, \bibinfo {author}
  {\bibfnamefont {D.}~\bibnamefont {Wu}}, \bibinfo {author} {\bibfnamefont
  {C.}~\bibnamefont {Lu}},\ and\ \bibinfo {author} {\bibfnamefont
  {C.}~\bibnamefont {Petrovic}},\ }\bibfield  {title} {\bibinfo {title}
  {Observation of multiple metastable states induced by electric pulses in the
  hysteresis temperature range of {{1$T$-TaS$_2$}}},\ }\href
  {https://doi.org/10.1103/PhysRevB.99.045102} {\bibfield  {journal} {\bibinfo
  {journal} {Phys. Rev. B}\ }\textbf {\bibinfo {volume} {99}},\ \bibinfo
  {pages} {045102} (\bibinfo {year} {2019})}\BibitemShut {NoStop}%
\bibitem [{\citenamefont {Givens}\ and\ \citenamefont
  {Fredericks}(1977)}]{givensThermalExpansionOp1977}%
  \BibitemOpen
  \bibfield  {author} {\bibinfo {author} {\bibfnamefont {F.~L.}\ \bibnamefont
  {Givens}}\ and\ \bibinfo {author} {\bibfnamefont {G.~E.}\ \bibnamefont
  {Fredericks}},\ }\bibfield  {title} {\bibinfo {title} {Thermal expansion op
  {{NbSe$_2$}} and {{TaS$_2$}}},\ }\href
  {https://doi.org/10.1016/0022-3697(77)90008-7} {\bibfield  {journal}
  {\bibinfo  {journal} {J. Phys. Chem. Solids}\ }\textbf {\bibinfo {volume}
  {38}},\ \bibinfo {pages} {1363} (\bibinfo {year} {1977})}\BibitemShut
  {NoStop}%
\bibitem [{\citenamefont {Yang}\ \emph {et~al.}(2024)\citenamefont {Yang},
  \citenamefont {Lee}, \citenamefont {Bang}, \citenamefont {Kim}, \citenamefont
  {Wulferding}, \citenamefont {Lee},\ and\ \citenamefont
  {Cho}}]{yangOriginDistinctInsulating2024}%
  \BibitemOpen
  \bibfield  {author} {\bibinfo {author} {\bibfnamefont {H.}~\bibnamefont
  {Yang}}, \bibinfo {author} {\bibfnamefont {B.}~\bibnamefont {Lee}}, \bibinfo
  {author} {\bibfnamefont {J.}~\bibnamefont {Bang}}, \bibinfo {author}
  {\bibfnamefont {S.}~\bibnamefont {Kim}}, \bibinfo {author} {\bibfnamefont
  {D.}~\bibnamefont {Wulferding}}, \bibinfo {author} {\bibfnamefont {S.-H.}\
  \bibnamefont {Lee}},\ and\ \bibinfo {author} {\bibfnamefont {D.}~\bibnamefont
  {Cho}},\ }\bibfield  {title} {\bibinfo {title} {Origin of {{Distinct
  Insulating Domains}} in the {{Layered Charge Density Wave Material
  1$T$-TaS$_2$}}},\ }\href {https://doi.org/10.1002/advs.202401348} {\bibfield
  {journal} {\bibinfo  {journal} {Adv. Sci.}\ }\textbf {\bibinfo {volume}
  {11}},\ \bibinfo {pages} {2401348} (\bibinfo {year} {2024})}\BibitemShut
  {NoStop}%
\bibitem [{\citenamefont {Zong}\ \emph {et~al.}(2018)\citenamefont {Zong},
  \citenamefont {Shen}, \citenamefont {Kogar}, \citenamefont {Ye},
  \citenamefont {Marks}, \citenamefont {Chowdhury}, \citenamefont {Rohwer},
  \citenamefont {Freelon}, \citenamefont {Weathersby}, \citenamefont {Li},
  \citenamefont {Yang}, \citenamefont {Checkelsky}, \citenamefont {Wang},\ and\
  \citenamefont {Gedik}}]{zongUltrafastManipulationMirror2018}%
  \BibitemOpen
  \bibfield  {author} {\bibinfo {author} {\bibfnamefont {A.}~\bibnamefont
  {Zong}}, \bibinfo {author} {\bibfnamefont {X.}~\bibnamefont {Shen}}, \bibinfo
  {author} {\bibfnamefont {A.}~\bibnamefont {Kogar}}, \bibinfo {author}
  {\bibfnamefont {L.}~\bibnamefont {Ye}}, \bibinfo {author} {\bibfnamefont
  {C.}~\bibnamefont {Marks}}, \bibinfo {author} {\bibfnamefont
  {D.}~\bibnamefont {Chowdhury}}, \bibinfo {author} {\bibfnamefont
  {T.}~\bibnamefont {Rohwer}}, \bibinfo {author} {\bibfnamefont
  {B.}~\bibnamefont {Freelon}}, \bibinfo {author} {\bibfnamefont
  {S.}~\bibnamefont {Weathersby}}, \bibinfo {author} {\bibfnamefont
  {R.}~\bibnamefont {Li}}, \bibinfo {author} {\bibfnamefont {J.}~\bibnamefont
  {Yang}}, \bibinfo {author} {\bibfnamefont {J.}~\bibnamefont {Checkelsky}},
  \bibinfo {author} {\bibfnamefont {X.}~\bibnamefont {Wang}},\ and\ \bibinfo
  {author} {\bibfnamefont {N.}~\bibnamefont {Gedik}},\ }\bibfield  {title}
  {\bibinfo {title} {Ultrafast manipulation of mirror domain walls in a charge
  density wave},\ }\href {https://doi.org/10.1126/sciadv.aau5501} {\bibfield
  {journal} {\bibinfo  {journal} {Science Advances}\ }\textbf {\bibinfo
  {volume} {4}},\ \bibinfo {pages} {eaau5501} (\bibinfo {year}
  {2018})}\BibitemShut {NoStop}%
\bibitem [{\citenamefont {Zhu}\ \emph {et~al.}(2015)\citenamefont {Zhu},
  \citenamefont {Cao}, \citenamefont {Zhang}, \citenamefont {Jia},
  \citenamefont {Guo}, \citenamefont {Yang}, \citenamefont {Zhu}, \citenamefont
  {Zhang}, \citenamefont {Plummer},\ and\ \citenamefont
  {Guo}}]{zhuHighResolutionElectron2015}%
  \BibitemOpen
  \bibfield  {author} {\bibinfo {author} {\bibfnamefont {X.}~\bibnamefont
  {Zhu}}, \bibinfo {author} {\bibfnamefont {Y.}~\bibnamefont {Cao}}, \bibinfo
  {author} {\bibfnamefont {S.}~\bibnamefont {Zhang}}, \bibinfo {author}
  {\bibfnamefont {X.}~\bibnamefont {Jia}}, \bibinfo {author} {\bibfnamefont
  {Q.}~\bibnamefont {Guo}}, \bibinfo {author} {\bibfnamefont {F.}~\bibnamefont
  {Yang}}, \bibinfo {author} {\bibfnamefont {L.}~\bibnamefont {Zhu}}, \bibinfo
  {author} {\bibfnamefont {J.}~\bibnamefont {Zhang}}, \bibinfo {author}
  {\bibfnamefont {E.~W.}\ \bibnamefont {Plummer}},\ and\ \bibinfo {author}
  {\bibfnamefont {J.}~\bibnamefont {Guo}},\ }\bibfield  {title} {\bibinfo
  {title} {High resolution electron energy loss spectroscopy with
  two-dimensional energy and momentum mapping},\ }\href
  {https://doi.org/10.1063/1.4928215} {\bibfield  {journal} {\bibinfo
  {journal} {Rev. Sci. Instrum.}\ }\textbf {\bibinfo {volume} {86}},\ \bibinfo
  {pages} {083902} (\bibinfo {year} {2015})}\BibitemShut {NoStop}%
\end{thebibliography}%

\end{document}